\documentclass[10pt,letterpaper]{article}
\usepackage[top=1in, bottom=1in, left=1in, right=1in]{geometry}

\usepackage{amsmath}
\usepackage{amsfonts}
\usepackage{amssymb}
\usepackage{multicol}
  \usepackage{makecell}
  
\usepackage{mathrsfs}
\usepackage{array}
\usepackage{tabu}
\usepackage{multirow}

\usepackage{color}
\usepackage[colorlinks]{hyperref}
\usepackage{hypernat}

\usepackage{graphicx}
\usepackage{subfigure} 

\newcommand\pd[2]{\mathchoice%
	{\frac{\partial #1}{\partial #2}}%
	{\partial #1/\partial #2}%
	{\partial #1/\partial #2}%
	{\partial #1/\partial #2}%
}

\newcommand\pdd[2]{\mathchoice%
	{\frac{\partial^2 #1}{\partial #2^2}}%
	{\partial^2 #1/\partial #2^2}%
	{\partial^2 #1/\partial #2^2}%
	{\partial^2 #1/\partial #2^2}%
}

\title{Small-scale inhibiting characteristics of \\ residual and solution filtering\protect\thanks{This work has been approved for public release (AFRL Public Affairs Clearance Number AFRL-2022-1521).}}

\begin{document}

  \author{
      Ayaboe K. Edoh\thanks{corresponding author: ayaboe.edoh@jacobs.com} \quad 
      Timothy P. Gallagher$^\ddag$ \quad Venkateswaran Sankaran$^*$    \\
           {\small \dag  {\itshape Jacobs Engineering, Inc. -- \itshape
       Air Force Research Laboratory, Edwards, California}} \\
           {\small \ddag  {\itshape Aerospace Systems Directorate -- \itshape
      Air Force Research Laboratory, Edwards, California}}  \\
                 {\small *  {\itshape Aerospace Systems Directorate -- \itshape
       Air Force Research Laboratory, Wright-Patterson, Ohio}}           
  }
\date{}
\maketitle

\graphicspath{{./figures_static/}}

\abstract{
Residual and solution filtering procedures are studied with respect to inhibiting the accumulation of small-scale (i.e., high wavenumber) content. 
Assessing each method in terms of an ``equivalent residual equation" reveals fundamental differences in their behaviors, such as how the underlying solution can be constrained to a target filter width.
The residual filtering (RF) approach paired with a dissipative filter kernel is shown to restrict scale generation in the fluid equations via dispersive effects; meanwhile, solution filtering (SF) -- and artificial dissipation (AD), by extension -- operates through dissipative mechanisms and actively attenuates high wavenumber content.
Discrete filters (i.e., the Top-hat and implicit Tangent schemes) are analyzed in terms of their response characteristics and their associated effects on reducing small-scale activity when paired with the RF versus SF approaches.
Linear theoretical assessments (e.g., von Neumann analysis) are shown to successfully characterize the fundamental behaviors of the methods in non-linear settings, as observed through canonical test cases based on 1D viscous Burgers, 2D Euler, 3D Navier-Stokes equations.
}

\section{Introduction} \label{sec:1}

Filtering methods serve to reduce unwanted information from signals, including: biases in the underlying system, noise that may arise due to the collection or generation of data streams, or the stochastic nature of the original source.
Each of these general uses aims to remove problematic content and finds an analogy in the context of computational fluid dynamics (CFD). 
\begin{enumerate}
\item Signal bias: discretizing the continuous equations induces numerical errors due to finite precision, with increasing impact on accuracy towards smaller scales \cite{Lele:1992}.
\item Noise from data generation: the creation of finer scales due to non-linearities in the equations can produce aliasing errors on a finite grid which can affect both accuracy and stability \cite{Moin:1997,Ghosal:1996,Kennedy:2008}.
\item Stochasticity: the coupled effects of numerical errors in the presence of non-linearities without regularization further challenges achieving convergent results and tractable flow characterizations (e.g., in turbulence \cite{Pope:2000,Sagaut:2009}).
\end{enumerate}
The desire to mitigate such issues leads to the use of filtering schemes in CFD. 
The corresponding flow equations feature active spectral redistribution of energy yet may exhibit a lack of natural regularization on under-resolved grids for dealing with errors that may appear in the numerical field.
Notable examples include high Reynolds number Large-Eddy Simulations (LES) of the Navier-Stokes equations, which often require model-based damping to maintain stability and accuracy.

Filters are typically incorporated for attenuation purposes targetting high-frequency (i.e., small time scales) or high-wavenumber (i.e., small spatial scales) components of the solution.
However, the way in which the filtering is employed as part of the computation plays an important role with respect to the numerical properties that influence accuracy and robustness.
The current work looks to analyze the nuances associated with two prominent filtering approaches for fluid equations: the residual filtering and solution filtering implementations.

The residual filtering (RF) methodology involves applying filter operators on the spatial derivatives and source terms within a time-marching procedure.
Uses of RF include increasing the stable time step size of explicit time integrators and accelerating the convergence of iterative methods. 
For example, Jameson and Baker \cite{Jameson:1983} propose smoothing (i.e., filtering) the residuals in order to accommodate arbitrarily large time steps, attributing the increased stability to the larger support of the effective difference operator. 
Meanwhile, Haelterman \emph{et al.} \cite{Haelterman:2010} optimize both explicit and implicit residual smoothers relative to a RK-multigrid framework in order to accelerate iterative performance, with the authors noting the importance of clustering high frequency modes relative to the damping properties of the RK-multigrid method.
Although the various contexts of RF are often considered independently, the notion of filtering the equation's time derivative (i.e., the residual) can be described most generally from the perspective of differential preconditioning \cite{Turkel:1993}.
More specifically, employing spatial attenuation to the time increment of the solution not only modifies its temporal advancement (whether in physical time or in an iterative pseudo time setting) but also introduces wavenumber-dependent biases such that the evolution of different scales can be manipulated.
The ability to tune filters relative to the governing equations as well as to the spectral content of the signal can therefore provide the necessary flexibility towards increasing the stability limit of explicit integrators (e.g., by artificially shrinking the eigenvalues of the semi-discrete system relative to a given time step size) and  steady-state convergence acceleration (e.g., by focusing the iterative procedure on scales in need of further convergence). 

Rather than applying a filter to the residual, one can also apply it directly to the solution variables as an intermediate post-processing step.
This is referred to as solution filtering (SF), originally introduced by Shuman \cite{Shuman:1957} for smoothing erroneous high-wavenumber components in meteorological computations. 
Meanwhile, the identification of aliasing errors by Phillips \cite{Phillips:1959} as a principal source of numerical instability for non-linear problems on under-resolved grids has further motivated the use of SF.
Periodically removing spectral content below a prescribed cut-off in the solution works to de-aliasing results and increase non-linear robustness of the solution.
The use of SF has since expanded.
Visbal and Gaitonde \cite{Visbal:2001, Visbal:2002} use high-order solution filtering to enhance the solution quality and robustness of high-order numerical methods for aeroacoustics on cuvilinear meshes.
Meanwhile Kennedy and Carpenter \cite{Kennedy:1994} demonstrate the use of high-order
solution filtering to suppress high-wavenumber noise generated by high-order explicit schemes. 
This filtering method is used for direct numerical simulations of turbulent combustion \cite{Chen:2011}, where the high-wavenumber content is strongly coupled to all wavenumbers due to the non-linearities in the problem.
The fundamental link of solution filtering to artificial dissipation (AD) methods \cite{Yee:1999, Asthana:2015, Edoh:2018, Shu:2021} -- wherein SF may be viewed as a predictor-corrector implementation of the latter -- furthermore extends its applicability to other situations such as shock capturing \cite{Bogey:2009,Wissink:2018}.

Understanding the fundamental mechanisms behind the application of residual and solution filtering methods is essential for their proper use.
Bogey and Bailly \cite{Bogey:2007} and Ranocha \emph{et al.} \cite{Ranocha:2018a} compare the performance of RF and SF for the linear advection equation, noting fundamental differences in dispersion and dissipation properties.
In this current work, we utilize both theoretical analysis and numerical examples to further identify the characteristics of residual and solution filtering methods when applied to the time evolution of non-linear 
equations of increasing complexity (i.e., the 1D viscous Burgers, 2D Euler, and 3D Navier-Stokes equations).
We show that both methods are viable for mitigating the presence of undesirable content in the systems; however, the mechanisms by which the content is altered is different for residual versus solution filtering.
For example, RF operates on the time derivative of the solution and can be said to impart a scale-dependent mollification of the original dynamics; consequently, the technique does not provide direct dissipation for actively removing unwanted content that may otherwise accumulate over time (e.g., spurious numerical errors). 
On the other hand, SF removes unwanted content by directly regularizing the solution; yet, it is sensitive to admitting excess numerical damping based on the choice of filter operator. 
Such fundamental differences may have implications for the respective use of residual and solution filtering in solving non-linear problems, including impacts on turbulence calculations and explicitly-filtered LES \cite{Oefelein:2023}.
The performance of artificial dissipation (AD) -- which can be seen as a special rendition of solution filtering \cite{Edoh:2018} -- is also touched upon.

The paper is organized as follows: Section \ref{sec:2} assesses the residual and solution filtering methods theoretical by considering their application to a prototypical linear scalar advection-diffusion equation, where the numerical dissipation and dispersion terms arising from the respective formulations are identified and von Neumann analysis is used to further study the overall scheme impact; Section \ref{sec:3} then presents non-linear numerical calculations that demonstrate and confirm the various conclusions gathered from the previous linear analysis; finally, Section \ref{sec:4} provides summarizing remarks on the explicit filtering implementations and proposes lines of future work.

\section{Methodology}  \label{sec:2}

The residual and solution filtering methods are considered relative to a general evolution equation written in residual form,
\begin{equation}
\pd{Q}{t} = \mathcal{R}(Q) \ , \label{eq:1}
\end{equation}
where $\mathcal{R}(Q)$ is the residual operator made up of spatial derivatives and source terms of the solution quantity $Q$.
To enable the theoretical analysis in this section, the one-dimensional linear advection-diffusion equation for the scalar $u$ is taken as a prototypical example for the target Navier-Stokes system:
\begin{equation}
\mathcal{R}(u) = - a \pd{u}{x} + \nu \pdd{u}{x} \ . \label{eq:2}
\end{equation}
The filter kernels in question are purely dissipative and are expressed in terms of a preserving and an attenuating operator in space,
\begin{equation}
\begin{aligned}
\mathcal{G}_{\text{fil}} = \mathcal{I} + \mathcal{D}_{\text{fil}} = 1 + \sum_{k\geq 1} \epsilon_{2k} (\Delta x)^{2k} \delta^{2k}_x \ , \label{eq:3}
\end{aligned}
\end{equation}
where $\delta_x^{m}$ is the discrete $m$-th derivative in the $x$-direction.
Here, we further assume that $(\delta^2_x)^k = \delta_x^{2k} = \partial_x^{2k} + O(\Delta x)^2$; note that the the non-divided difference, $\delta^{2k} = (\Delta x^{2k}) \delta^{2k}_x$, may be used interchangeably for convenience.
Specific stencils associated with the filters are responsible for defining the coefficients $\epsilon_{2k}$.
Their impact on the different methods are theoretically investigated, with this linear perspective providing fundamental understandings into the methods' non-linear behaviors.

\subsection{Overview of Filtering Implementations} \label{sec:2a}

The filtering techniques considered in this work are presented side by side in Table \ref{table:1} in order to facilitate comparisons.
In addition to RF and SF, artificial dissipation (AD) is also studied.
Each of the methods are decomposed into three basic steps: the forming of the residual, the application of the filter, and the temporal integration of the system.
The order in which each of these is executed varies depending on the method.

Here, the filtered quantity is denoted by an overbar, $\bar{\phi} = \mathcal{G}\{\phi\}$.
In general, both filtering residual and solution filtering work as a post-processing of the input $\phi$, where we take $\phi = \mathcal{R}_{\text{o}}(Q)$ in the case of residual filtering and $\phi = Q^{n+1,*}$ in the case of solution filtering.
The filter operation can furthermore be re-expressed as a temporal correction step, 
\begin{equation}
\begin{aligned}
\bar{\phi} &= \phi + \sum_k \epsilon_{2k}(\Delta x)^{2k}\delta_x^{2k}\phi =  \phi + (\Delta t) \cdot \sum_k \epsilon_{2k} \lvert \lambda' \rvert (\Delta x)^{2k-1}\delta_x^{2k}\phi \ , \label{eq:4}
\end{aligned}
\end{equation}
where $\lambda'$ is a parameter with velocity units.

As proposed in previous work \cite{Edoh:2018}, temporal consistency in Equation \ref{eq:4} is achievable by re-scaling the original filter coefficients by a CFL related parameter, $\epsilon_{2k}' = \mu \epsilon_{2k}$ where $\mu = \min\{CFL,1 \}$\footnote{Alternatively, Lamballais \emph{et al} \cite{Lamballais:2021} suggest a scaling based on the viscous dynamics, where $\mu = 1 - e^{-\text{VNN}}$ and VNN $= \nu \Delta t/(\Delta x^2)$. This technique also recovers time consistency of the solution filtering procedure. A convective-based rendition in this form would also be possible by substituting the VNN by the CFL number, although the analogy to traditional artificial dissipation schemes is lost.} and $CFL = \lvert \lambda \rvert \Delta t/\Delta x$.
Doing so implies that $\lambda' = \min\{ \Delta x/\Delta t,a\}$ is related to a physical velocity scale for CFLs below unity -- which is consistent with what is typically done for artificial dissipation methods. 
On the other hand, omitting such a re-scaling suggests a discretization-based velocity scale of $\lambda' = (\Delta x/\Delta t)$.
In the case of solution filtering, this purely grid-based definition will increase the amount of numerical dissipation imparted onto the solution as $(\Delta t) \to 0$.
The CFL-based re-scaling of solution filtering avoids the ambiguity of determining how often to filter the solution by enforcing a physical-based frequency of damping.
Unless otherwise noted, this temporally-consistent re-scaling formulation is assumed henceforth for solution filtering and is referred to as SF-r.

\begin{table}
\center
\caption{Algorithmic description of the filtering approaches.}
\label{Tab:FilterAlgorithm}
\begin{tabular}{l | l | l}
Artificial Dissipation (AD) & Residual Filtering (RF)  & Solution Filtering (SF) \\ \hline  & & \\
 1. Form the base residual, $\mathcal{R}_{\text{o}}(Q)$  & 1. Form the base residual, $\mathcal{R}_{\text{o}}(Q)$ & 1. Form the base residual, $\mathcal{R}_{\text{o}}(Q)$ \\ & & \\
\makecell[l]{2. Form the AD residual: \\ $\mathcal{R}_{\text{AD}}(Q) = \sum_k \epsilon_{2k}\lvert \lambda \rvert (\Delta x)^{2k-1} \delta_x^{2k}Q$} &
\makecell[l]{2. Filter the base residual: \\ $\mathcal{G}\{\mathcal{R}_{\text{o}}\} = \mathcal{R}_{\text{o}} + \sum_k \epsilon_{2k}(\Delta x)^{2k}\delta^{2k}_x\mathcal{R}_{\text{o}}$} &
\makecell[l]{2. Integrate the residual in time: \\ $\partial_t Q = \mathcal{R}_{\text{o}}(Q), \ Q^n \to Q^{n+1,*}$} \\ & & \\
\makecell[l]{3. Integrate the combined \\ residuals in time: \\ $\partial_t Q = \mathcal{R}_{\text{o}}(Q)+ \mathcal{R}_{\text{AD}}(Q), \ Q^{n} \to Q^{n+1}$} &
\makecell[l]{3. Integrate the filtered residual in time: \\ $\partial_t Q = \mathcal{G}_{\text{fil}}\{\mathcal{R}_{\text{o}} \}, \ Q^n \to Q^{n+1}$} &
\makecell[l]{3. Filter the solution: \\ $Q^{n+1} = \mathcal{G}_{\text{fil}}\{Q^{n+1,*}\}$ }
\end{tabular}\label{table:1}
\end{table}

In order to assess the nuances between residual and solution filtering, we apply these relative to the discretized linear advection-diffusion equation (see Equation \ref{eq:2}).
Assuming a forward Euler integration scheme for simplicity and then constructing the methods using the steps outlined in Table \ref{table:1}, one gets
\begin{equation}
\frac{u^{n+1} - u^n}{\Delta t} = - a \delta_x u + \nu \delta_x^2 u^n + \mathcal{R}_{\text{addi}}(u^n) \ ,
\end{equation}
where the induced spatial artificial terms are
\begin{eqnarray}
\mathcal{R}_{\text{addi,AD}}(u^n) &=& \underbrace{\lvert \lambda' \rvert  \sum_k \epsilon_{2k} (\Delta x)^{2k-1}\delta_x^{2k} u^n}_{\color{red}I}  \label{eq:7}  \\ \nonumber \\
\mathcal{R}_{\text{addi,RF}}(u^n) &=& -  \underbrace{(\Delta t) \cdot a \lvert \lambda' \rvert \sum_k \epsilon_{2k}  (\Delta x)^{2k-1}\delta_x^{2k+1} u^n}_{\color{blue} II} 
+ \underbrace{(\Delta t) \cdot  \nu \lvert \lambda' \rvert \sum_k \epsilon_{2k}  (\Delta x)^{2k-1}\delta_x^{2k+2} u^n}_{\color{cyan} III} \label{eq:6} \\ \nonumber \\
\mathcal{R}_{\text{addi,SF}}(u^n) &=& \underbrace{\lvert \lambda' \rvert  \sum_k \epsilon_{2k} (\Delta x)^{2k-1}\delta_x^{2k} u^n}_{\color{red} I} 
 \label{eq:8} \\ 
 &&  - \underbrace{(\Delta t) \cdot a \lvert \lambda' \rvert  \sum_k \epsilon_{2k} (\Delta x)^{2k-1}\delta_x^{2k+1} u^n}_{\color{blue} II} +  \underbrace{(\Delta t) \cdot \nu \lvert \lambda' \rvert  \sum_k \epsilon_{2k} (\Delta x)^{2k-1}\delta_x^{2k+2} u^n}_{\color{cyan} III}    \nonumber
\end{eqnarray}
The above represent ``equivalent residual equations" (ERE) that help to highlight the additional mechanisms (i.e., $R_{\text{addi}}$) implied by each filtering method\footnote{The ERE yields overall spatial operators at play. This is in contrast to modified equation analysis (see Appendix \ref{app:a}) which takes into account the implied coupling effects with the time integration method, expressed in a PDE form.}. 

The artificial dissipation rendition includes the dissipative part of the filter as the additional contribution (see term ${\color{red} I}$ in Equation \ref{eq:7}) and therefore has an active mechanism for attenuating high wavenumber content.
The added artificial term can interact with the temporal method and result in secondary dispersive effects -- for details see the modified equation analysis provided in Appendix \ref{sec:a}.

In the case of residual filtering, two additional terms are induced (see Equation \ref{eq:6}).
The term ${\color{blue} II}$ is dispersive and is responsible for affecting the phase characteristics of the solution.
Meanwhile, the term ${\color{cyan} III}$ is related to the physical diffusion and can be said \emph{a priori} to be anti-dissipative, assuming the original filter is stable such that $\lvert \hat{\mathcal{G}} \rvert \le 1$\footnote{Note that the filter coefficients $\epsilon_{2k}$ of a stable filter are such that the Fourier response of the attenuating part of the filter is negative semi-definite, $\mathcal{F}\{\sum_k \epsilon_{2k}\delta^{2k}\} \le 0$. Multiplying this sum by $\delta^2$ then automatically makes the contribution anti-dissipative.}.
Whether the induced numerical anti-diffusion simply diminishes the effects of physical diffusion versus actively instigates a numerical instability will depend on the specific filter coefficients.
As will be shown later on, filters that feature negative regions in their responses (i.e., $\hat{\mathcal{G}} \in [-1,1]$), such as the Top-hat kernel, will yield unstable algorithms when residual filtering is applied to physical diffusion terms.
Importantly, the ERE in Equation \ref{eq:6} contains no dissipative terms and therefore the RF procedure has no mechanism to remove unwanted high wavenumber content that may be introduced into the solution.

The ERE associated with solution filtering (see Equation \ref{eq:8}) shows yet another combination of the induced mechanisms.
It features an AD term ${\color{red} I}$ that accounts for attenuation, but it also includes the dispersive and anti-diffusive terms ${\color{blue} II}$ and ${\color{cyan} III}$ from the ERE of residual filtering.
At first sight, this would suggest that solution filtering will induce both dispersive and anti-diffusion effects; however, this is misleading.
As explained in previous work \cite{Edoh:2018}, the terms ${\color{blue} II}$ and ${\color{cyan} III}$ in Equation \ref{eq:8} are induced by coupling the filtering with the time integrated result, and they represent cancellations to the secondary effects induced by the attenuating effect of term ${\color{red} I}$.
Note, for example, that the modified equation analysis of AD presented in Appendix \ref{sec:a} reveals analogous contributions to terms ${\color{blue} II}$ and ${\color{cyan} III}$ in Equation \ref{eq:8}, but of opposite sign.
Therefore the impact of solution filtering is confirmed to be uniquely dissipative, as expected.

\subsection{Characteristics of Discrete Filter Stencils} \label{sec:2b}

The previous section establishes the theoretical behavior of general filter operators as applied to the residual or to the solution.
Specifying the form of the filter kernel further reveals how these theoretical behaviors can manifest.
To this end, we focus here on the Top-hat filters as well as the implicit Tangent filters by Raymond \cite{Raymond:1988}.
Their responses are provided in Table \ref{table:2}, where $k_\Delta$ is the characteristic cut-off wavenumber. 
Both filtering schemes can be translated into discrete stencils of the form
\begin{eqnarray}
\overbrace{a_o \bar{\phi}_i+ \sum_{\ell \ge 1} a_\ell (\bar{\phi}_{i+\ell} + \bar{\phi}_{i-\ell})}^{\left[1 + \sum_{\ell \geq 1} \epsilon_{2\ell} (\Delta x)^{2\ell} \delta^{2\ell}_x\right]\{\bar{\phi}\}} = \overbrace{b_o\phi_i + \sum_{r \ge 1} b_r (\phi_{i+r} + \phi_{i-r})}^{ \left[1 + \sum_{r\geq 1} \epsilon_{2r} (\Delta x)^{2r} \delta^{2r}_x\right]\{\phi\}} \ .
\end{eqnarray}

For the Top-hat filter, the corresponding discrete stencil coefficients $b_r$ are recovered by first defining the filter as a convolution, $\bar{\phi}_i = \frac{1}{\Delta}\int_{x_i - \Delta/2}^{x_i + \Delta/2} [\phi(y)] dy$,
and then approximating the integral with a quadrature rule.
In the following, we assume a composite trapezoidal rule for simplicity.
The characteristic length of the Top-hat (i.e., the filter width, $\Delta$) defines the averaging window and induces a cut-off wavenumber $k_\Delta = (2\pi/\Delta)$.

Meanwhile, the discrete form of the Tangent filter is most easily expressed in terms of the $\epsilon_{2\ell/2r}$ coefficients.
These can be derived by first re-writing the Tangent response in terms of powers of $[\sin^2\left(k\Delta x/2\right)]$, and then employing the fact that $\mathcal{F}\{\delta^{2m}\} = [\cos(k\Delta x) - 2]^m = \left[-4\sin^2\left(k\Delta x/2\right)\right]^m$.
The Tangent filter -- unlike the Top-hat filter -- does not have a clear characteristic length in physical space.
{Usually, such discrete filters are instead defined in terms of their spectral response, such that $\hat{\mathcal{G}}(k_\Delta) = 0.5$ \cite{Lund:1997}.
Here, however, we choose the definition $\hat{\mathcal{G}}(k_\Delta) = 0.99$ in order to define the resolved modes as those that are well preserved by the filter. 

\begin{table}
\center
\caption{Filter definitions.}
\label{Tab:FilterResponses}
\begin{tabular}{c | c | c}
Filter & Response ($\hat{\mathcal{G}}$) & \makecell[c]{Characteristic  \\wavenumber ($k_\Delta$)}  \\ \hline  & \\
Top-hat & \makecell[l]{$\hat{\mathcal{G}}_{\text{TH}}(k) = \text{sinc}(\mu_\Delta \cdot k\Delta x)$ \\ \quad with $\mu_\Delta = \frac{\Delta}{(2\Delta x)}$} & $k_\Delta = \frac{2\pi}{\Delta}$  \\ & & \\
Tangent \cite{Raymond:1988} & \makecell[l]{$\hat{\mathcal{G}}_{\text{Tan}}(k) = \left[1 + \mu_\Delta \tan^{2R}\left(\frac{k\Delta x}{2} \right) \right]^{-1}$ \\ \quad with $\mu_\Delta = \frac{1/\hat{\mathcal{G}}(k_\Delta)-1}{\tan^{2R}\left(\frac{k_\Delta \Delta x}{2} \right)}$} &  $\hat{\mathcal{G}}(k_\Delta) = 0.99$ \\ & & \\
Spectral Sharp & \multicolumn{1}{l |}{ $\hat{\mathcal{G}}_{\text{sharp}}(k) = \left\{\begin{array}{r l} 1, & \text{if} \ k \le k_\Delta \\ 0, & \text{otherwise}  \end{array} \right.$} & $k_\Delta = \frac{2\pi}{\Delta}$
\end{tabular}\label{table:2}
\end{table}

\begin{figure} [h!]
 \centering
 \subfigure[]{ \label{fig:1a}
 \includegraphics[width=0.45\textwidth]{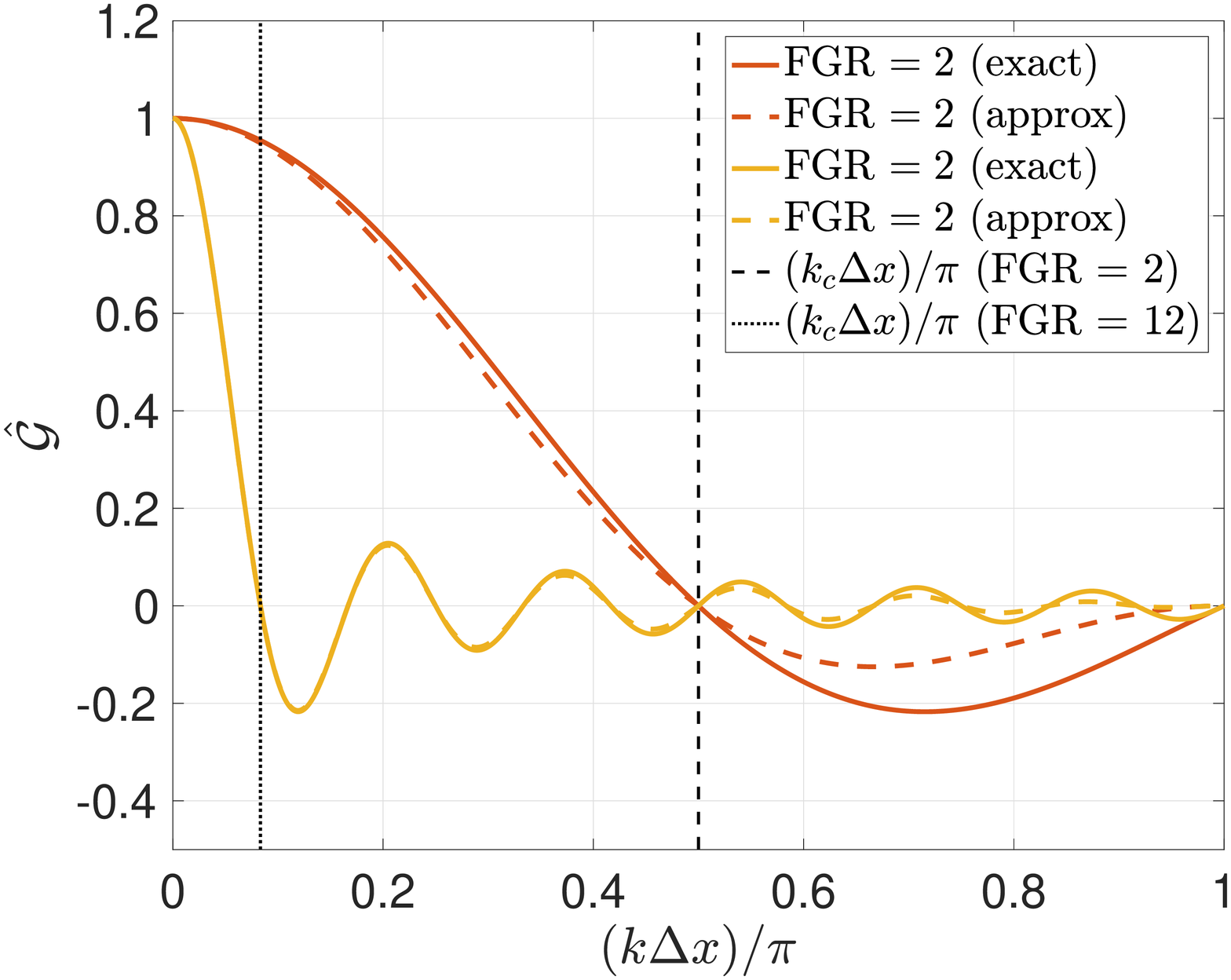}}
  \subfigure[]{ \label{fig:1b}
 \includegraphics[width=0.45\textwidth]{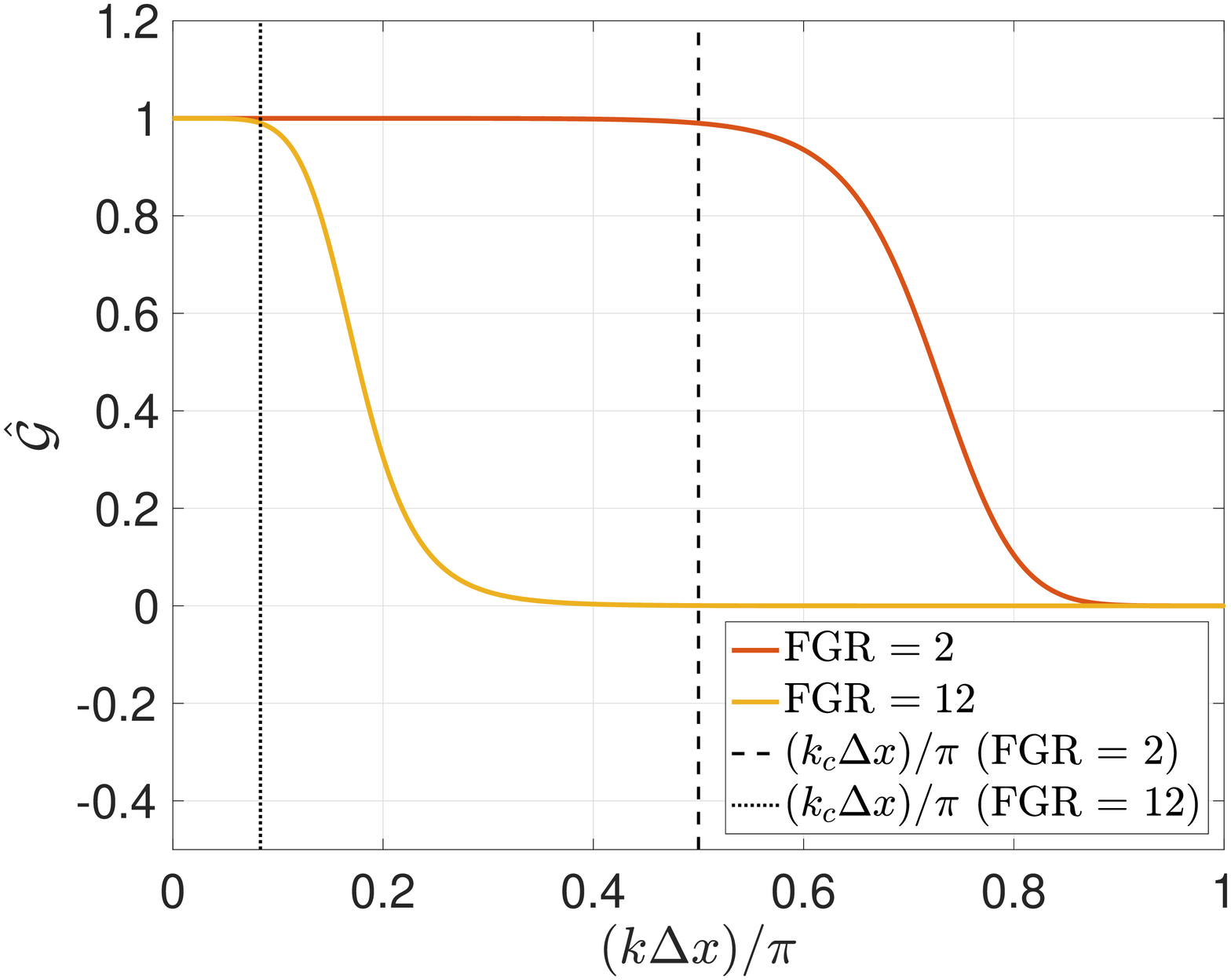}}
 \caption{Growth factor $\hat{\mathcal{G}}$ filter responses versus normalized wavenumber at different FGR (denoted by $k_c$): a) Top-hat filter, and b) sixth-order Tangent filter. }
  \label{fig:1}
\end{figure}

Figure \ref{fig:1} shows the responses of the filters as tuned to different filter-to-grid ratios (FGR), defined as
\begin{equation}
FGR = \frac{\pi}{(k_\Delta \Delta x)} \ .
\end{equation}
Figure \ref{fig:1a} plots the exact response of the Top-hat filter (i.e., the sinc function).
Note that the first zero is found at the cut-off wavenumber that is associated with the FGR, $k_\Delta  = \frac{\pi}{(FGR \cdot \Delta x)}$, 
and subsequent crossings occur at integer harmonics of $k_\Delta$ .
In the case where FGR $> 1$, the response features negative values.
The negative response of the filter flips the original magnitude of a given signal -- which, in the case of a sinusoidal basis, would be interpreted as shifting the signal completely out of phase.
In the next section, this property of the response will be shown to affect the stability and phase characteristics of the residual filtering methods.
Also included in Figure \ref{fig:1a} are the responses associated with the discrete stencil that arises from approximating the convolution integral for a top-hat filter with a composite trapezoidal quadrature rule.
Note that the approximation replicates the exact response at lower wavenumbers and only differs slightly in amplitude at the high wavenumbers; the proper zero-crossings, however, are maintained.
This approximation will be utilized henceforth.

Meanwhile Figure \ref{fig:1b} shows Fourier response for sixth-order Tangent filters as tuned to different FGR per the current $\hat{\mathcal{G}}(k_\Delta x) = 0.99$ definition.
Compared to the Top-hat filter, we note that the Tangent response is positive semi-definite and only reaches zero at the Nyquist wavenumber, $(k\Delta x) = \pi$.
In addition, the attenuation decreases monotonically and shows strong scale separation -- a characteristic that increases with the order of the stencil.

\subsection{Von Neumann Analysis} \label{sec:2c}

The performance of the filters with respect to their different implementations is now analyzed in terms of the dissipation and dispersion characteristics of the overall discretization of the equations.
Von Neumann analysis (VNA) is employed, and the procedure is motivated from a semi-discrete Ordinary Differential Equation (ODE) perspective in order to highlight the fundamental trends between the different methods.

Assuming a discrete Fourier representation for the solution variable,
\begin{equation}
u_{i+m} = \sum_k \hat{u}_k \cdot e^{\imath k(x_i+ \Delta x_m)} \ ,
\end{equation}
then the prototypical linear advection-diffusion equation of Equation \ref{eq:2} is re-written in a semi-discrete form that describes the evolution each Fourier mode,
\begin{equation}
\pd{\hat{u}_k}{t} = \underbrace{\left[- a\cdot k_{\text{conv}} + \nu \cdot k_{\text{diff}}\right]}_{\beta_{\text{o}}} \hat{u}_k \ . \label{eq:17}
\end{equation}
The modified wavenumbers, $k_{\text{conv}}$ and $k_{\text{diff}}$, quantify the effect of discrete difference stencils relative to exact differentiation.
For example, explicit central stencils on a uniform grid for the convection and diffusion terms yield modified wavenumbers
\begin{eqnarray}
\begin{aligned}
\mathcal{F}\left\{\delta_x u_i = \sum_{r\ge1} \frac{c_r}{(\Delta x)} (u_{i+r} - u_{i-r})\right\} & \to  \sum_k \hat{u}_k \cdot \overbrace{\left[\sum_{r \ge 1}\imath \cdot \frac{2c_r \sin(rk\Delta x)}{(\Delta x)}\right]}^{k_{\text{conv}}} e^{\imath kx_i} \\
\mathcal{F}\left\{\delta^2_x u_i = \sum_{r\ge0} \frac{d_r}{(\Delta x)^2} (u_{i+r} + u_{i-r})\right\} & \to  \sum_k \hat{u}_k \cdot \overbrace{\left[\sum_{r\ge 0} \frac{2d_r \cos(rk\Delta x)}{(\Delta x)^2}\right]}^{k_{\text{diff}}} e^{\imath kx_i}
\end{aligned}
\end{eqnarray}
and the analogous exact analytical differentiations relative to the Fourier basis gives the references
\begin{eqnarray}
\mathcal{F}\{\partial_x u \} \ \to \ \sum_k \hat{u} \cdot \overbrace{\left[\imath \cdot k \right]}^{k_{\text{conv,ex}}} e^{\imath k x_i} \quad \text{and} \quad
\mathcal{F}\{\partial^2_x u \} \ \to \ \sum_k \hat{u} \cdot \overbrace{\left[-k^2 \right]}^{k_{\text{diff,ex}}} e^{\imath k  x_i}  \ .
\end{eqnarray}
Considering the ODE that results from the semi-discretization in Equation \ref{eq:17}, the complex-valued amplification factor $\hat{\mathcal{G}}(k)$ that describes the evolution of each mode is
\begin{equation}
\hat{\mathcal{G}} = \lvert \hat{\mathcal{G}} \rvert e^{\imath \omega} \approx e^{(\beta \Delta t)} \quad \text{such that} \quad \hat{u}^{n+1} = \hat{\mathcal{G}} \cdot \hat{u}^{n} \ . \label{eq:20}
\end{equation}
The above is equivalent to the ODE characteristic polynomial evaluated with respect to the system eigenvalues $\beta$, which in turn are parametrized by the wavenumber dependent (e.g., $\beta = -a \cdot k_{\text{conv}}(k) + \nu \cdot k_{\text{diff}}$).
In Equation \ref{eq:20}, $\lvert \mathcal{G} \rvert$ corresponds to the growth factor (i.e., dissipation) of the method while $\omega$ characterizes its phase properties (i.e., dispersion) .
Assuming a Runge-Kutta (RK) time scheme, the characteristic polynomial is defined as \cite{Butcher:2003}
\begin{equation}
\hat{\mathcal{G}}_{\text{RK}}\left(\beta \Delta t \right) = 1 + (\beta \Delta t) \cdot {\bf b}^T\left[I - (\beta \Delta t) \cdot A \right]^{-1}{\bf 1} \ , \label{eq:21}
\end{equation}
where $[A,{\bf b}]$ are the RK coefficients arranged in Butcher Tableau form and $\beta(k)$ are the eigenvalues associated with the semi-discrete system.
The final amplification factors for the different filtering implementations are expressed succinctly as
\begin{equation}
\begin{aligned}
\underline{\text{Artificial Dissipation (AD)}}: \quad \hat{\mathcal{G}}_{\text{AD}} &= \hat{\mathcal{G}}_{\text{RK}}\left(\beta_{\text{o}}\Delta t + \beta_{\text{AD}}\Delta t \right), \ \text{with} \ \beta_{\text{AD}} = \frac{\lvert \lambda' \rvert}{(\Delta x)} \cdot \overbrace{\left[\hat{\mathcal{G}}_{\text{fil}} - 1\right]}^{\hat{\mathcal{D}}_{\text{fil}}} \\
\underline{\text{Residual Filtering (RF)}}: \quad \hat{\mathcal{G}}_{\text{RF}} &= \hat{\mathcal{G}}_{\text{RK}}\left(\hat{\mathcal{G}}_{\text{fil}} \cdot (\beta_{\text{o}} \Delta t) \right) \\
\underline{\text{Solution Filtering (SF)}}: \quad \hat{\mathcal{G}}_{\text{SF}} &= \hat{\mathcal{G}}_{\text{fil}} \cdot \hat{\mathcal{G}}_{\text{RK}}\left(\beta_{\text{o}}\Delta t \right)
\end{aligned}
\end{equation}
Note that the above presumes that solution filtering occurs at each new time step -- rather than after each stage of the RK method, which would more closely resemble an AD implementation.

\begin{figure} [h!]
 \centering
 \subfigure[]{ \label{fig:2a}
 \includegraphics[width=0.45\textwidth]{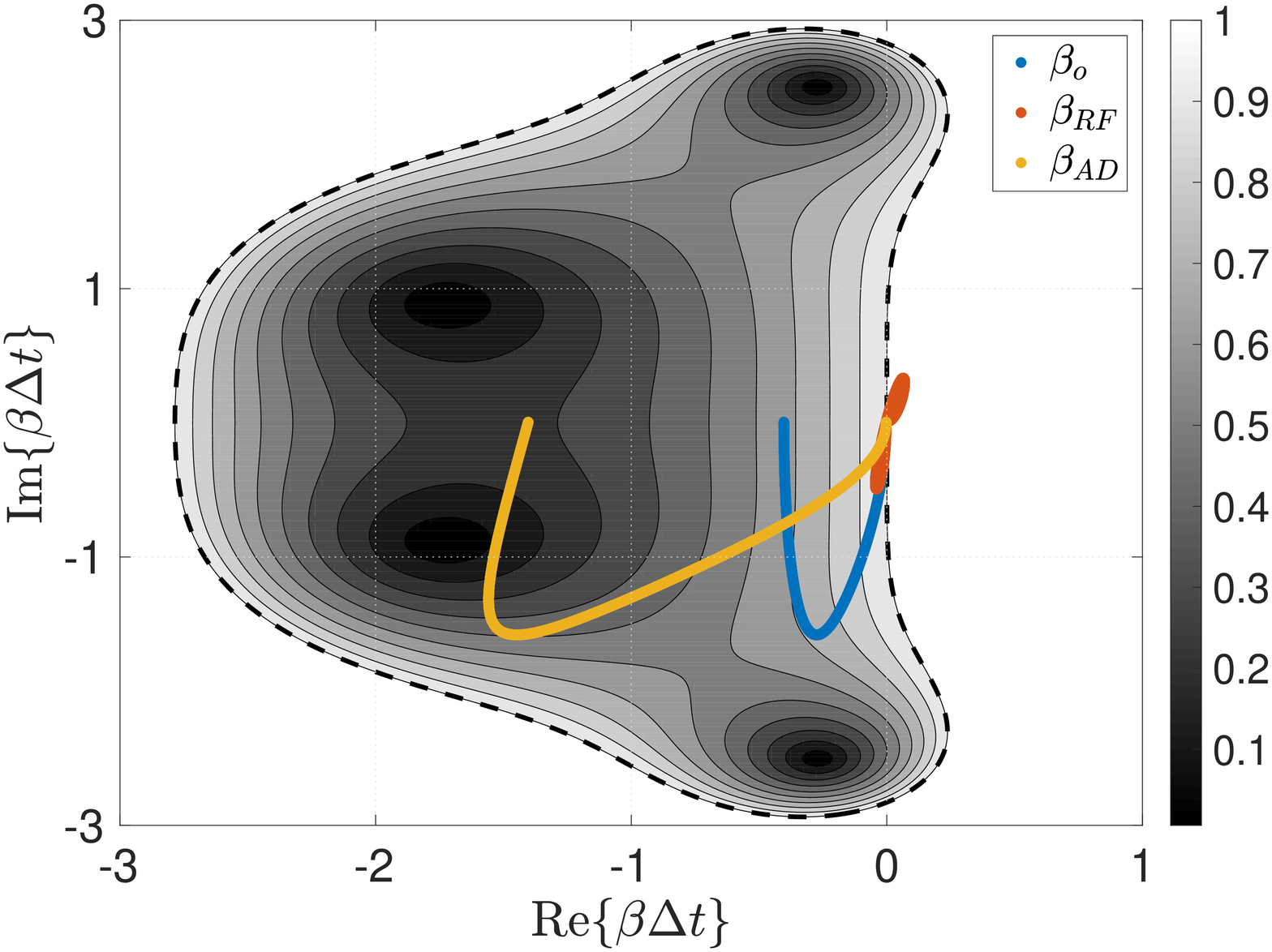}}
  \subfigure[]{ \label{fig:2b}
 \includegraphics[width=0.45\textwidth]{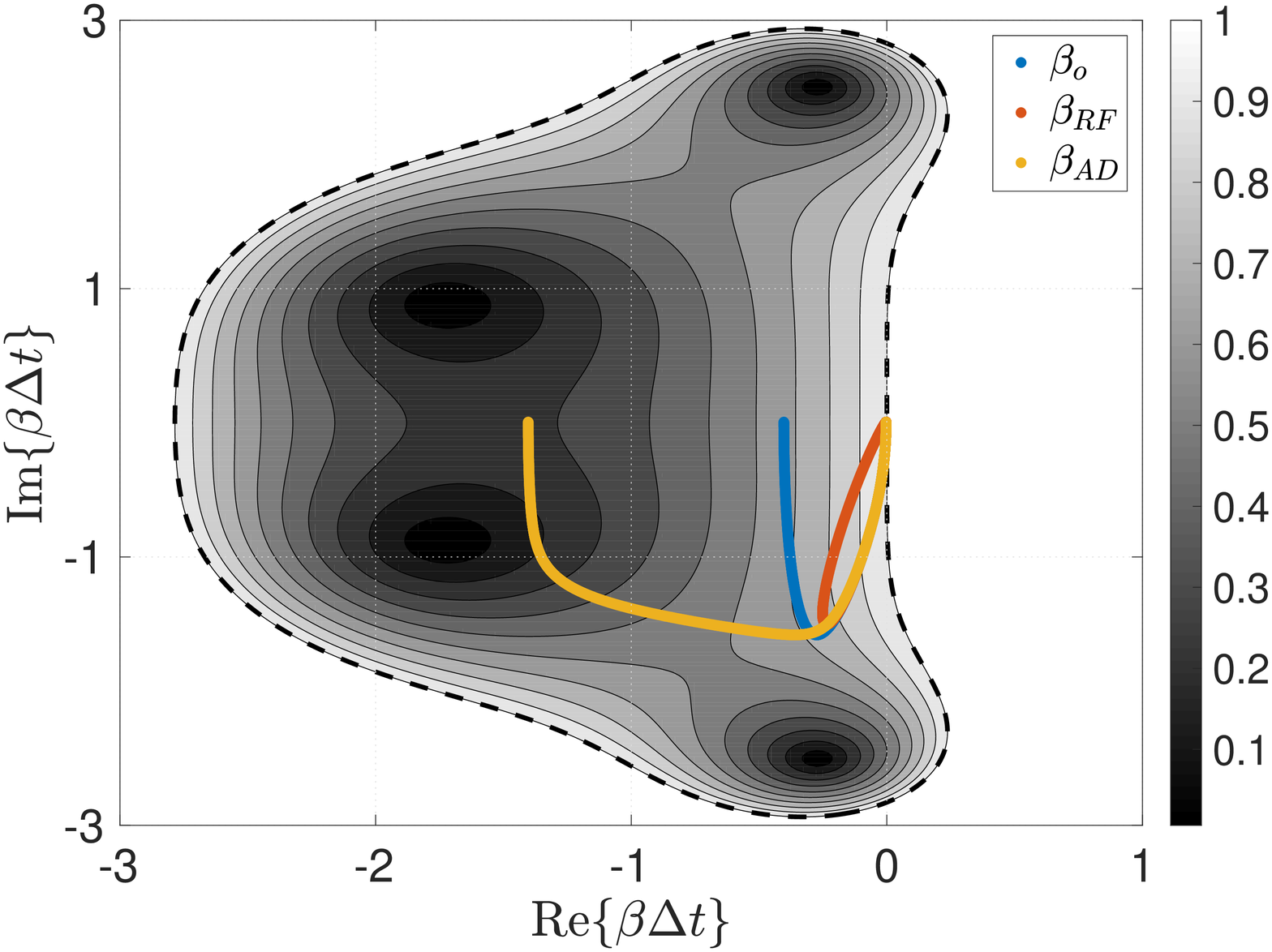}}
 \caption{Scatter of semi-discrete eigenvalues arising from the discretization of the advection-diffusion equation according to the different filtering methodologies set to FGR $= 2$ ($\beta_o$: no filtering, $\beta_{RF}$: residual filtering, $\beta_{AD}$: artificial dissipation), plotted on the magnitude of the integration method amplification factor ($\lvert \hat{\mathcal{G}}_{RK} \rvert$) for the classic fourth-order RK method: a) Top-hat filter and b) sixth-order Tangent filter.}
 \label{fig:2}
\end{figure}

Figures \ref{fig:2} shows the eigenvalues associated with the Fourier representation of the linear advection-diffusion equation (see Equation \ref{eq:17}) with and without filtering, wherein we consider a standard sixth-order central discretization of the convective term and a standard second order central stencil for the diffusion term.
The central convective terms are purely dispersive and thus induce imaginary components to the eigenvalues; meanwhile the diffusion terms add a stabilizing negative real component.
The relative magnitude of the real versus imaginary components of $\beta_{\text{o}}$ is dictated by a numerical Reynolds number -- which has been set to $Re_{\Delta x} = (|a|\Delta x )/\nu = 10$ for this example.
Meanwhile the spectral radius of the eigenvalues is characterized by a time step size associated with $CFL = (|a|\Delta t)/(\Delta x) = 1$.
Top-hat and Tangent filters with FGR $=2$ are considered, and the eigenvalues of the resulting semi-discrete systems are plotted over the magnitude contours of the classic fourth-order Runge-Kutta integration method ($\lvert \hat{\mathcal{G}}_{\text{RK}} \rvert$) in order to better highlight temporal-spatio coupling effects.
The impact of filtering on the dissipation and dispersion characteristics of the overall method is therefore fully contextualized -- at least for RF and AD (note: SF includes an additional post-processing effect that is not captured here). 
Studying contour plots (i.e. ``thumbprints") of the characteristic polynomial as parametrized by a set of complex eigenvalues then provides additional insight into how the time step size, the discretization scheme, or the filtering method will impact the eventual von Neumann analysis.

Overall, the AD implementation is shown to add a real component to the original eigenvalues and thus increases integration stiffness.
On the other hand, RF is seen to generally shrink the distribution of eigenvalues which confirms the use of residual filtering (also referred to as residual smoothing) for accommodating higher CFL limits. 
Meanwhile no direct commentary on SF is possible from this semi-discrete perspective, due to its post-processing nature.

The influence of the respective filter functions on the eigenvalues is also evident.
The current Tangent filter preserves more of the baseline eigenvalues, and this property is tied to the fact that the specific response preserves more content and features enhanced scale-discriminant damping compared to the Top-hat, which modifies a larger portion of the original eigenvalue distribution. 
Also important to note is that the residual filtering with the current Top-hat filter instigates de-stabilizing positive real eigenvalues to the system.
This is related to the negative portion of the Top-Hat filter's response function (see Figure \ref{fig:1a}) that interacts with the diffusion terms and thus changes them into anti-diffusion terms.

\begin{figure} [h!]
 \centering
 \subfigure[]{ \label{fig:4a}
  \includegraphics[width=0.45\textwidth]{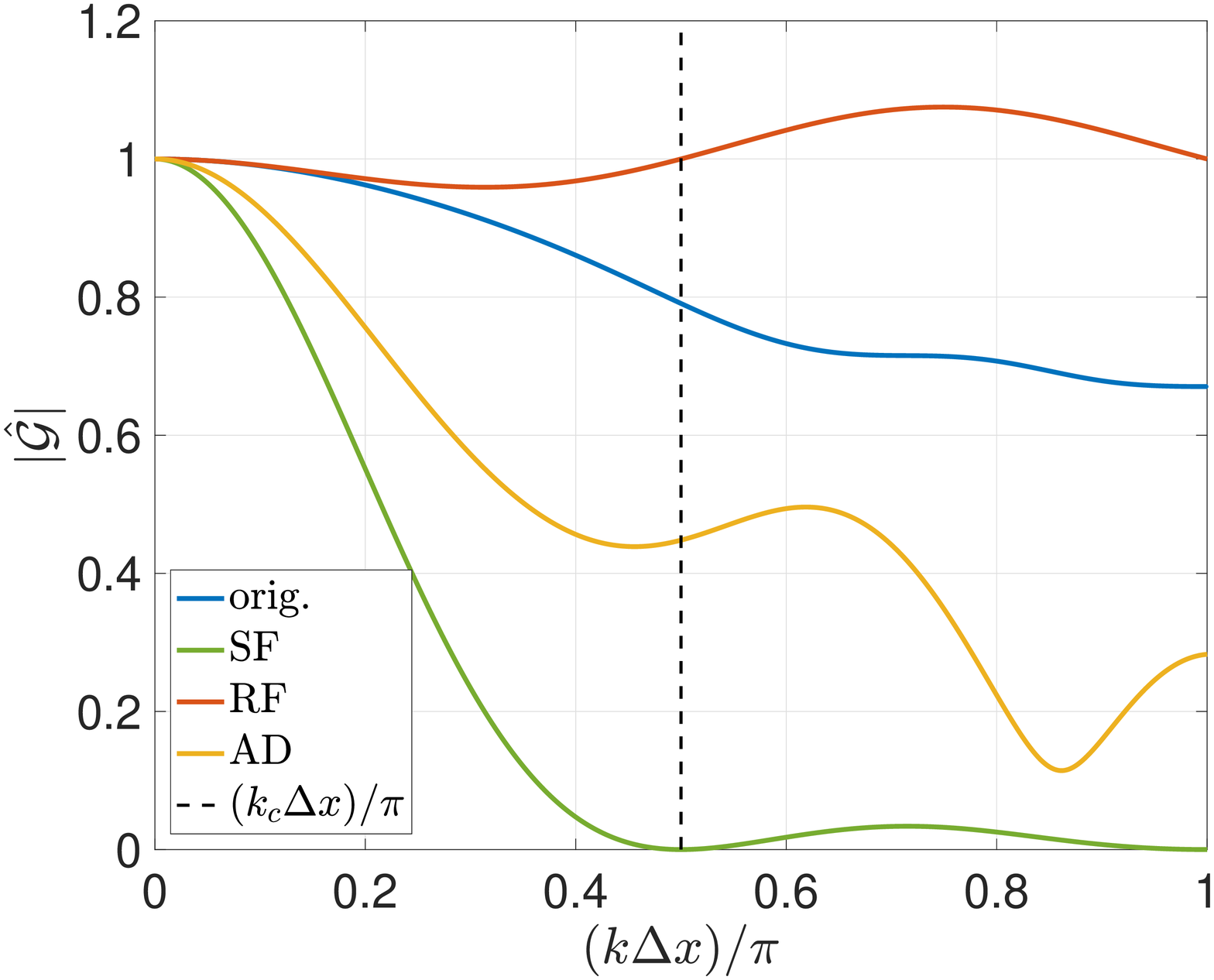}}
  \subfigure[]{ \label{fig:4b}
   \includegraphics[width=0.45\textwidth]{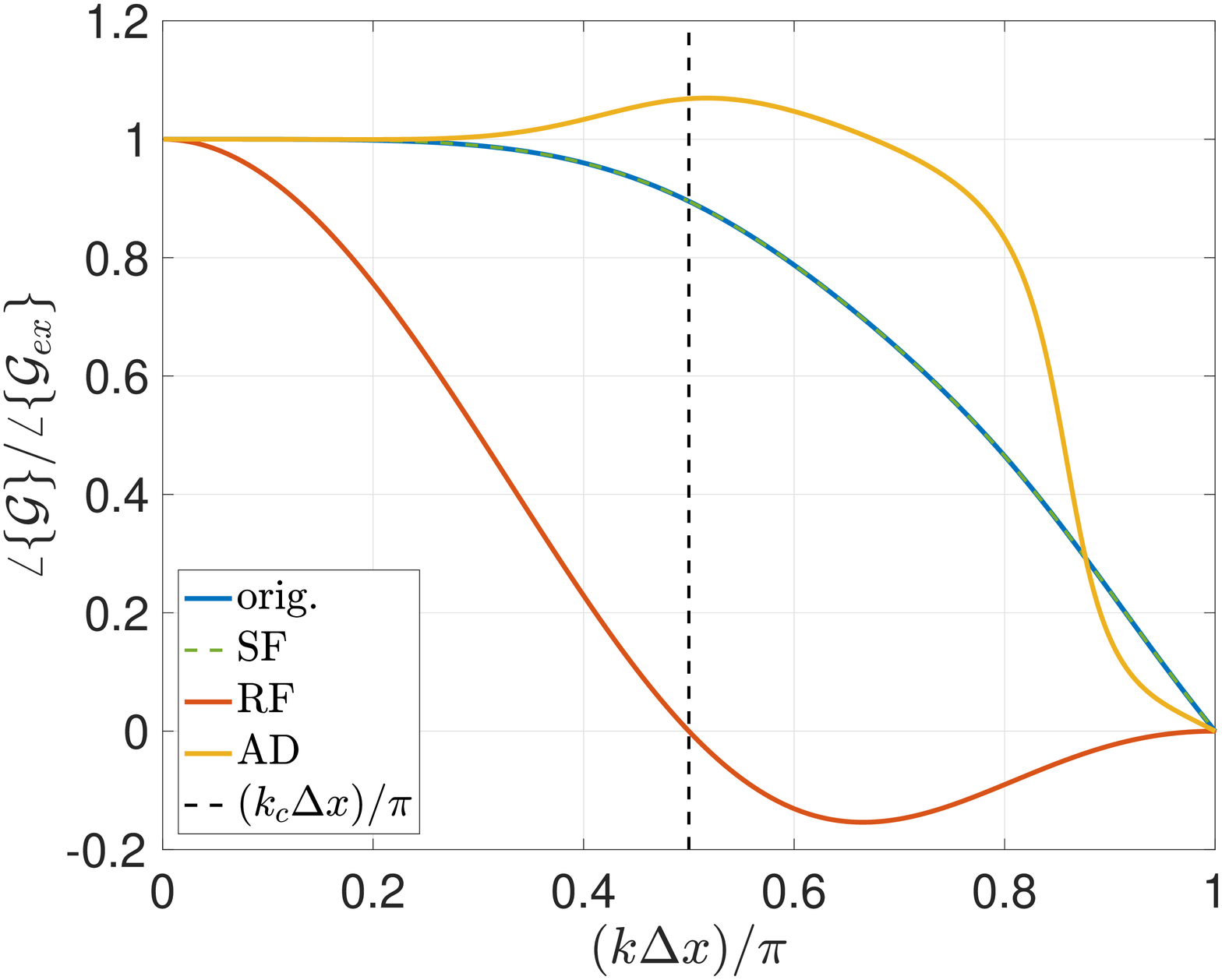}}
 \caption{Von Neumann analysis for filtering schemes tuned to FGR $= 2$ (denoted by $k_c$) with the Top-hat filter: a) growth factor and b) phase (normalized by the exact phase as calculated by assuming exact integration and a Fourier-spectral scheme in space).
 }
 \label{fig:4}
\vspace{5mm}
 \centering
 \subfigure[]{ \label{fig:5a}
   \includegraphics[width=0.45\textwidth]{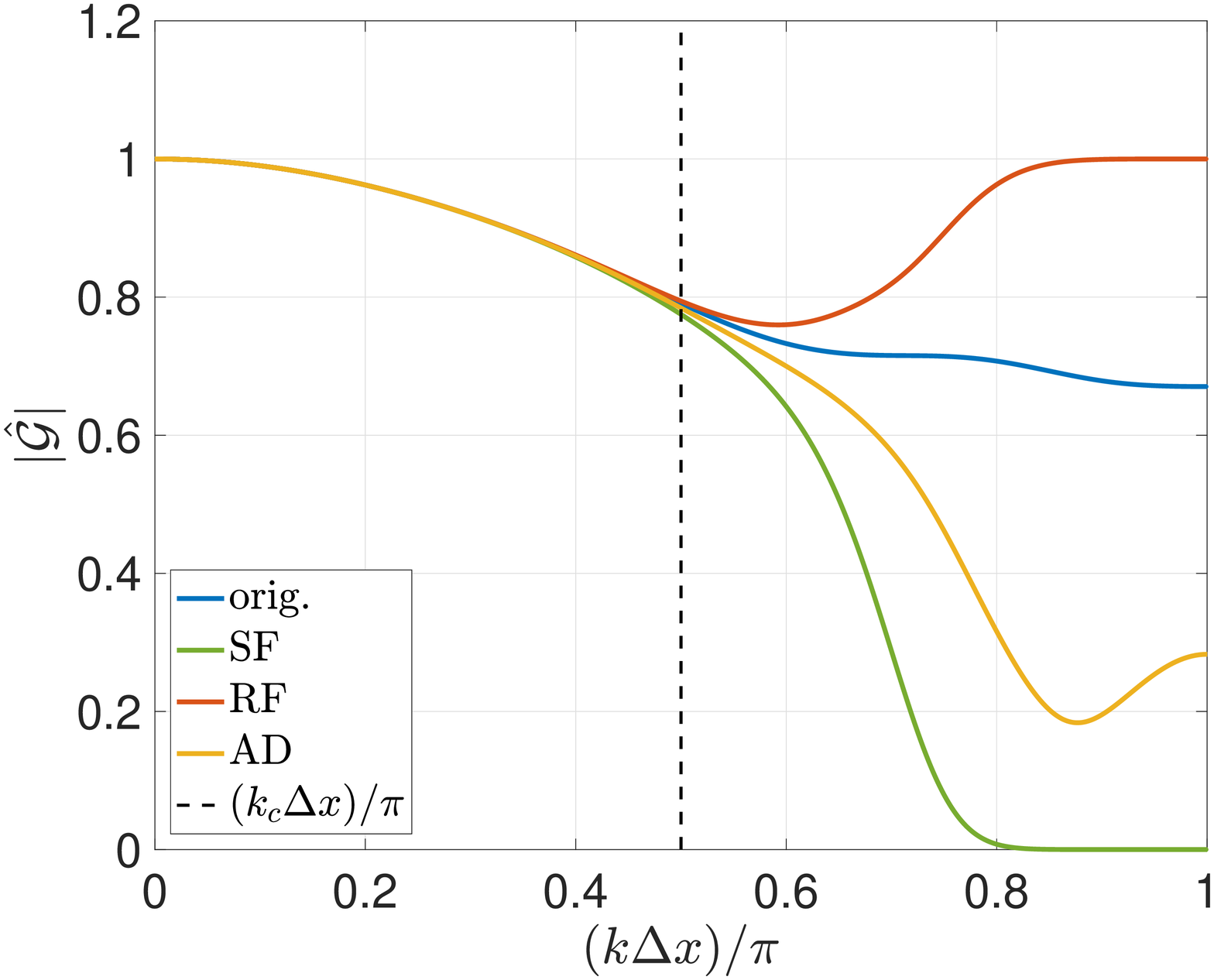}}
  \subfigure[]{ \label{fig:5b}
 \includegraphics[width=0.45\textwidth]{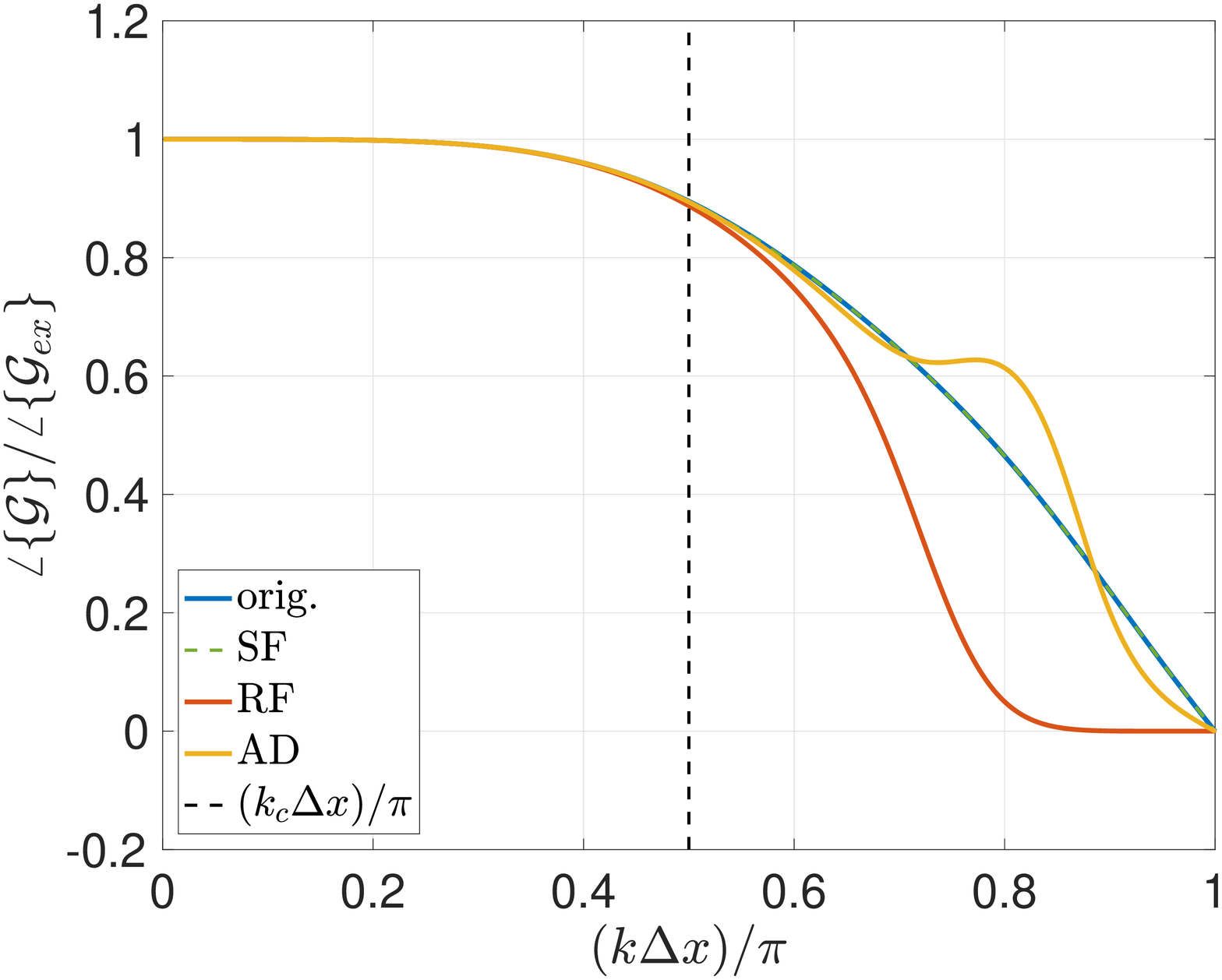}}
 \caption{Von Neumann analysis for filtering schemes tuned to FGR $= 2$ (denoted by $k_c$) with the sixth-order Tangent filter: a) growth factor and b) phase (normalized by the exact phase as calculated by assuming exact integration and a Fourier-spectral scheme in space). }
 \label{fig:5}
\end{figure}

The associated von Neumann analysis further details dissipation and dispersion properties of the different overall filtering methods as a function of wavenumber.
The stability plots follows from the one-to-one mapping that exists between the current semi-discrete eigenvalues and a wavenumber (see Equation \ref{eq:17}).
Results pertaining to both the Top-hat and sixth-order Tangent filters are provided in Figures \ref{fig:4} and \ref{fig:5}, respectively.
In general, one notes that the RF technique reduces the amount of damping and also decreases phase accuracy -- the extent of which depends on the filter response.
Meanwhile, the AD approach adds damping at the high wavenumbers and somewhat affects phase behavior.
This latter observation is a result of spatio-temporal coupling effects -- which, for example, are also responsible for producing non-monotonic damping at moderate CFL numbers.
The trends for RF and AD are both predictable by understanding how the filtering influences the semi-discrete eigenvalues and how these interact with the ODE thumbprints of the integration method (see Figure \ref{fig:2}).
SF, on the other hand, is an attenuating post processing method; it is seen to preserve the phase properties of the base scheme and to only add the filter dissipation to the base scheme.

The pairing of residual filtering with the Top-hat filter induces notable effects in terms of stability and phase accuracy.
In the instance where the Top-hat filter employs a FGR $ > 1$, then the associated filter response features negative values (see Figure \ref{fig:1a}).
In the case of residual filtering, this corresponds to the upper lobe of $\beta'_{RF}$ in Figure \ref{fig:2a}. 
With regard to dissipation, one notes that the same range of modes feature a growth factor above unity, therefore making the method unstable (see Figure \ref{fig:4a}).
This is a consequence of the physical diffusion terms being included in the residual filtering; therefore, a potential remedy is to exclude such terms from the RF procedure when employing filters that feature negative responses.
Meanwhile with regard to dispersion, one notes a range of modes (those associated with the negative response of the Top-hat filter) over which the phase direction is reversed (see Figure \ref{fig:4b}) -- this is consequential with respect to transport accuracy.
Furthermore, the zero crossings cause stationary phase and neutral dissipation at the respective wavenumbers.
Therefore one can expect content to become trapped at these spectral locations.
The Top-hat filter features zero crossings at $k_\Delta$ and its harmonics; thus, there is the potential for a pile up of spectral energy at multiple intermediate modes besides the Nyquist frequency when FGR $ > 1$.

Overall, the von Neumann responses for each filtering method are similar up until the point of significant filter roll-off, after which the nuances in dissipation and dispersion of each implementation become more evident.
These characterizations provide guidelines to the non-linear setting, which is explored in the following sections via numerical calculations.

\section{Results}  \label{sec:3}

In order to illustrate the different behaviors of the filtering implementations, we consider several non-linear test cases of increasing complexity on uniform periodic domains.
The previous linear analysis is shown to appropriately predict the behavior of the filtering implementations.
The strengths and weaknesses of the different approaches are explored by first studying non-linear cascading in the one-dimensional viscous Burgers equation.
Next, additional attention is placed on transport accuracy and noise mitigation with respect to an Euler computation of the two-dimensional isentropic vortex.
Finally, the Taylor-Green vortex, as governed by the three-dimensional Navier-Stokes system, is presented in order to highlight the impact of both dissipation and dispersion characteristics on the transient dynamics and also to study how each filtering method performs for varying FGR. 
In each instance, the Top-hat and Tangent filters are employed to further highlight the effects of their different response characteristics as analyzed in the previous section.
In the following cases, the AD method employs $| \lambda' |= \max \{| u | +c\}$ and the SF procedure is applied at each new $(n+1)$ time step with a CFL-based re-scaling such that $| \lambda' | = \min\{ \Delta x/\Delta t,\max \{| u | +c\}\}$.

\subsection{1D Viscous Burgers: Non-linear Cascade} \label{sec:3a}

A model problem based on the viscous Burgers equation is first used to highlight some of the differences between the filtering implementations, here focusing on their spectral characteristics in terms of enforcing a target filter width.
The non-linear dynamics of the Burgers equation result in the creation of smaller scales via non-linear interactions. 
The current intent of the filtering methods is thus to restrict the solution spectrum to be within the target filter width which is set to FGR $= 12$. 

\subsubsection{Initialization}

We seek the solution to the one-dimensional viscous Burgers equation, 
\begin{equation}
\pd{u}{t} + \frac{1}{2}\pd{u^2}{x} = \nu \pdd{u}{x}
\label{Eqn:Burgers}
\end{equation}
The domain is periodic domain, and the above equation is advanced with a second-order five-stage Low Dissipation Disperion Runge-Kutta (LDDRK) method \cite{Hu:1996,Stanescu:1998}. 
The convective term is represented in divergence form and is discretized with the central fourth-order, seven-point optimized  (CD04-7opt) Dispersion Relation Preserving (DRP) central stencil of Tam and Webb \cite{Tam:1993}.
Meanwhile the diffusion term employs a standard fourth-order central stencil.
The equations are integrated with a  time step corresponding to $CFL_{u_o} \approx 0.5$ which gives $\Delta t \approx 5.6 \times 10^{-5}$.
The initial solution imposes a prescribed energy spectrum via Fourier series \cite{San:2016}:
\begin{equation}
\begin{aligned}
\overbrace{u(t=0,x)}^{u_o} &= 2\sum_{n=0}^{N/2} \left[\Re\{\hat{u}(\kappa_n)\}  \cos\left(\frac{2\pi \kappa_nx}{L}\right) +\Im\{\hat{u}(\kappa_n)\}  \sin\left(\frac{2\pi \kappa_nx}{L}\right) \right] \\
\text{with}& \quad \hat{u}(\kappa) =
\sqrt{2E(\kappa)} e^{ \imath 2\pi \mathsf{U}_\kappa}, \ E(\kappa) = \left[\frac{2}{3\sqrt{\pi}}\left(\frac{2\pi\kappa_0}{L} \right)^{-5}\right] \left(\frac{2\pi \kappa}{L}\right)^4 e^{-(\kappa/\kappa_0)^2}
\end{aligned}
\end{equation}
In the above, $\kappa = \frac{2L}{2\pi}$ refers to the integral wavenumber.
Phase randomization is applied by choosing $\mathsf{U}_\kappa \in [0,1]$ from a uniform distribution.
The initial condition has a peak in the energy spectrum at wavenumber $\kappa_0 = 5$, which identifies an integral length scale $\ell \approx L/\kappa_o$.
The integral Reynolds number is set to $Re_\ell = \frac{(L/\kappa_o)}{\nu }\cdot \left(\frac{1}{L}\int u_o^2 dx\right)^{1/2}=1000$.
Employing $N_x = 8192$ degrees of freedom (DOF) therefore provides sufficient resolution of the viscous scales, $\eta/\Delta x \approx 4.5$ (based on Kolmogorov turbulence scaling approximations).
The large filter-to-grid ratio (FGR $=12$) to be considered, however, yields a resolution that lies within the non-viscous inertial range.

\subsubsection{Results}

\begin{figure} [h!]
 \centering
 \subfigure[]{ \label{fig:8a}
 \includegraphics[width=0.45\textwidth]{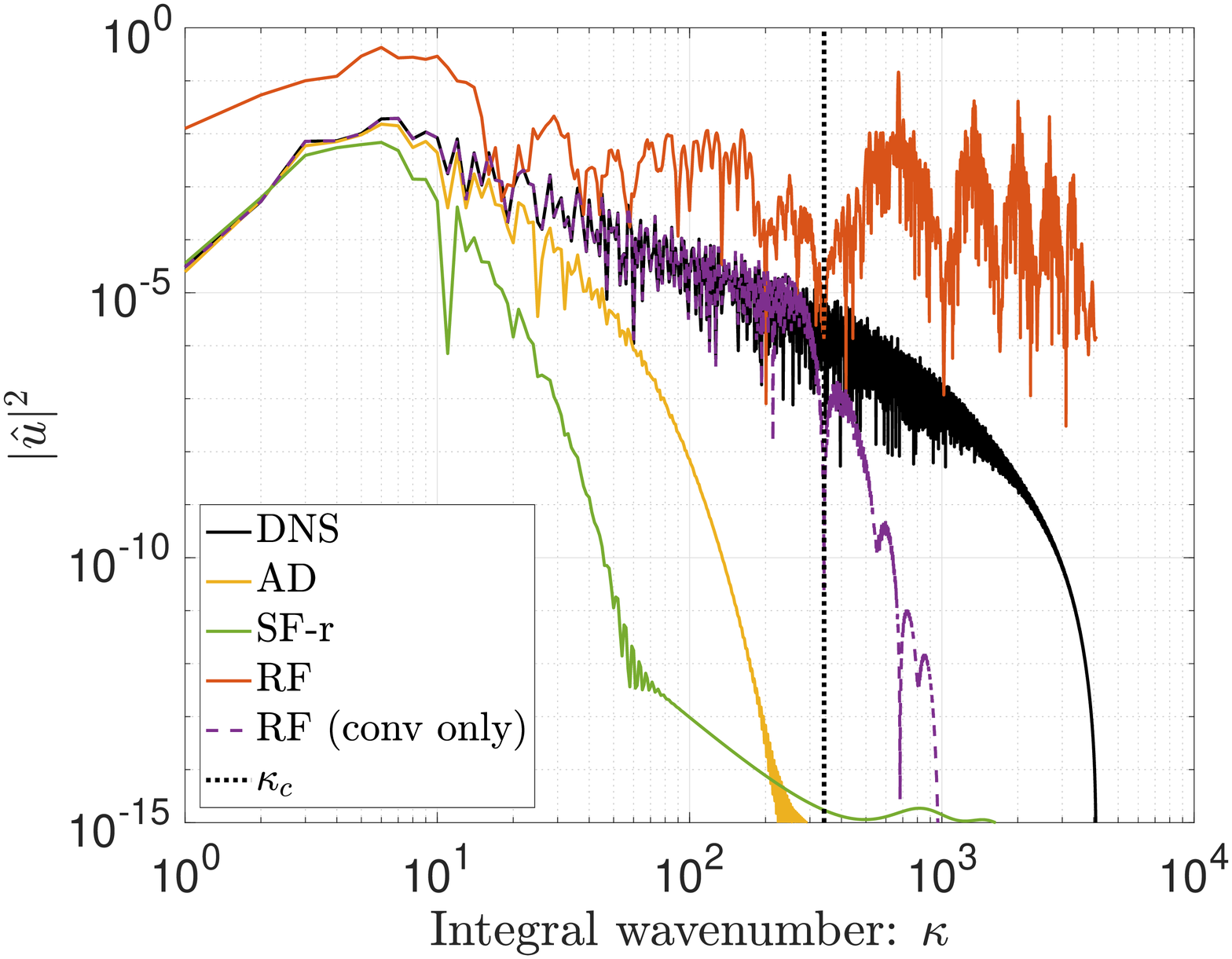}}
  \subfigure[]{ \label{fig:8b}
 \includegraphics[width=0.45\textwidth]{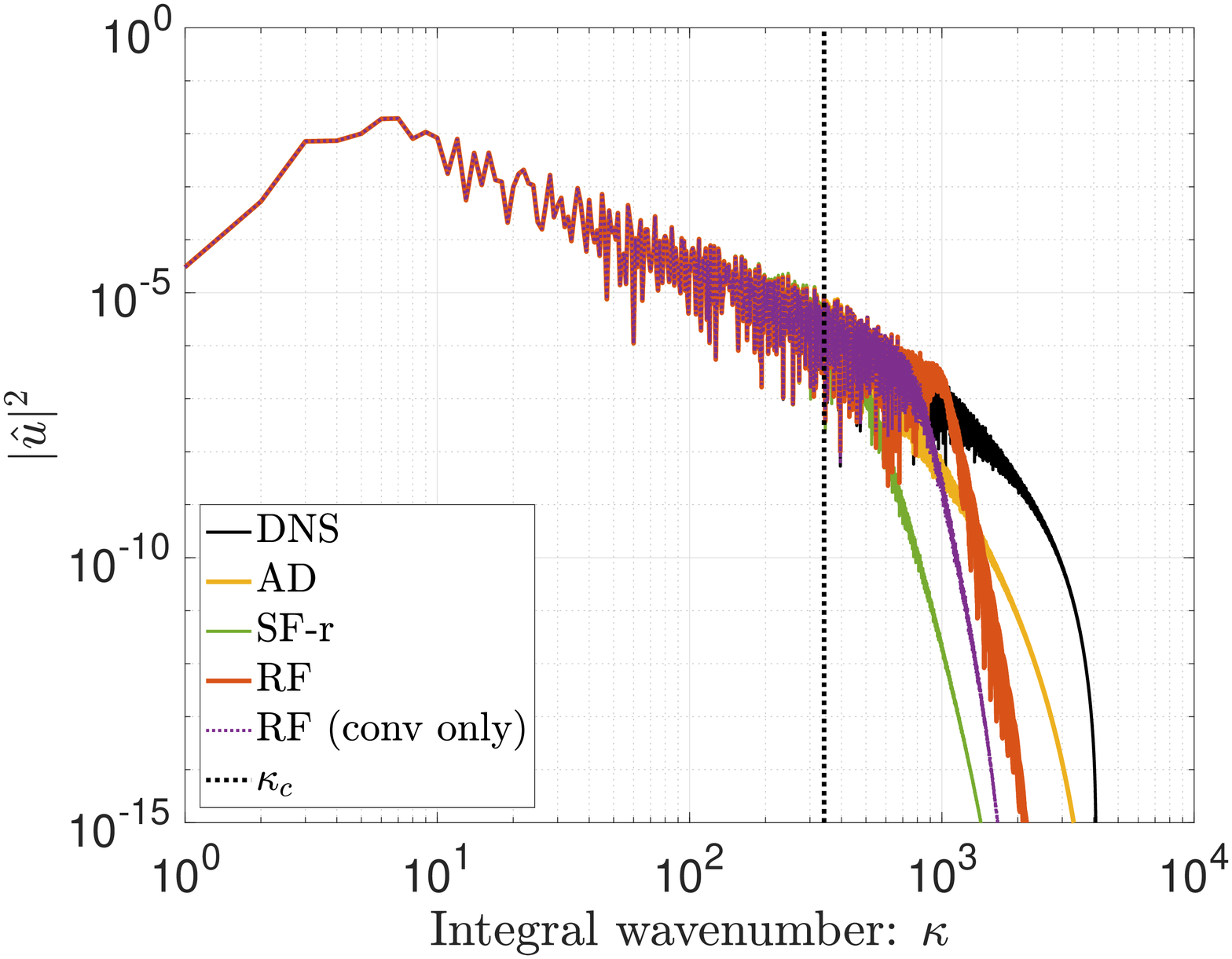}}
   \subfigure[]{ \label{fig:8c}
 \includegraphics[width=0.45\textwidth]{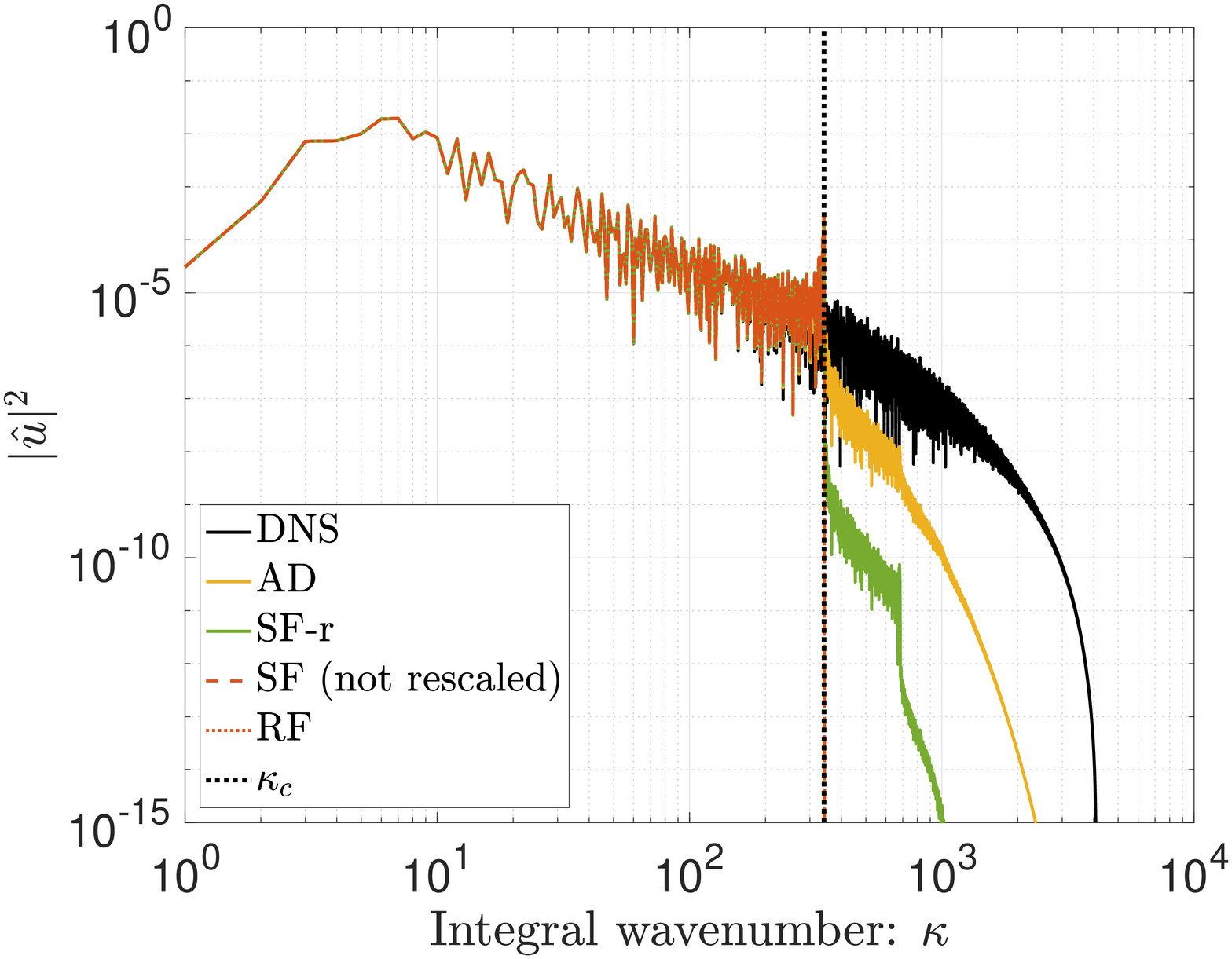}}
 \caption{Power spectral density of solution, $|\hat{u}|^2$, corresponding to ``turbulent" Burgers test case at $t = 0.0475$ (near peak dissipation rate), comparing different filtering implementations tuned to FGR $= 12$ (denoted by $\kappa_c$) relative to the full DNS resolution: a) Top-hat filter, b) sixth-order Tangent filter and c) spectral-sharp filter. }
 \label{fig:8}
\end{figure}

Figures \ref{fig:8} plots the solution spectra at $t = 0.0475$, which corresponds to the moment of maximum energy dissipation rate (i.e., the spectrum is at its ``fullest", prior to decay).
The performance of the different filtering methods can thus be compared in terms of their damping properties.
In general, one notes that solution filtering is the most dissipative approach, followed by artificial dissipation and then residual filtering.
The extent to which this attenuation impacts the scales above the filter size for the different implementations depends on the scale-discriminance of the filter in use.
For example, the Top-hat filter yields in stark differences between RF, SF, and AD; yet, such differences are less pronounced when using the scale-discriminant Tangent filter.

The role of the filter's scale-discriminance in spectral space is further highlighted in Figure \ref{fig:8c}, which shows the results for a spectrally sharp cut-off filter.
In this scenario, the associated solutions are nearly identical in terms of their sharp drop off at the target filter width, and they only show slight deviances at sub-filter scales.
Residual filtering and solution filtering without the CFL rescaling (SF-nr) yield identical results when paired with the sharp filter.
Nevertheless, each method is still distinct and would react differently depending on the choice of initial conditions, the introduction of noise, the time step size, etc. \footnote{In order to recover identical results with residual filtering versus solution filtering (SF-nr), one would need to employ a spectral sharp filter that is set at a filter width below the aliasing limit of the equations (i.e., enforce a de-aliased calculation); the initial condition would furthermore need to be contained within the prescribed filter width, and no other spurious noise (e.g., from boundaries, aliasing etc.) should be introduced since the methods react differently to such errors \cite{GallagherEdoh:2019a}.}

We note that the RF implementations tend to admit scales past the target cut-off $k_\Delta$ when utilizing filters with a smooth response.
The finite amplitude of the filter response beyond the prescribed cutoff admits this small-scale content.
As previously stated, RF does not have an active dissipation mechanism and therefore cannot damp the lingering high wavenumber components.
The influence of physical diffusion is furthermore reduced when the viscous terms are included in the residual filtering, therefore allowing for a greater accumulation of small-scale content.
In the case of the Top-hat filter -- which features a negative response over a range of wavenumbers (see Figure \ref{fig:1a}) -- one sees that including the diffusion term in the RF procedure introduces numerical instabilities in the corresponding scales.
Therefore RF filtering with a Top-hat kernel featuring FGR $ > 2$ should be performed only on the convective terms and should exclude the diffusion terms.

\subsection{2D Euler: Isentropic Vortex} \label{sec:3b}

Next, the performance of the filtering techniques is evaluated relative to mitigating numerical errors.
For the current transport-dominated isentropic vortex case, the compressible Euler equations are considered (see Equation \ref{eq:25}, neglecting the viscous terms, ${\bf E}_j^{(v)}$) while assuming a calorically perfect ideal gas.
The transport of the initial density profile is an exact solution to the governing inviscid equations; yet, numerical perturbations can trigger a non-linear cascade of spectral content that will cause the vortex to break up. 
Therefore the filtering formulations are inspected with regard to mitigating the manifestation of such numerical error effects, and the ability to maintain proper vortex coherence and transport.

\begin{eqnarray}
\partial_t \underbrace{\left[ \begin{array}{c} \rho \\ \rho u_i\\ \rho e_o \end{array} \right]}_{\bf Q}   =
- \partial_{x_j}\underbrace{\left[  \begin{array}{c} \rho u_j \\ \rho u_ju_i   \\  \rho u_je_o \end{array} \right]}_{{\bf E}_j^{(c)}}
- \partial_{x_j}\underbrace{\left[  \begin{array}{c} 0 \\  P\delta_{ij} \\  Pu_j \end{array} \right]}_{{\bf E}_j^{(p)}}
+ \partial_{x_j} \underbrace{\left[ \begin{array}{c}  0 \\ \tau_{ij} \\ \tau_{jk}u_k + q_j \end{array} \right]}_{{\bf E}_j^{(v)}} = \mathcal{R}_{\text{o}} \label{eq:25}
\end{eqnarray}
\begin{eqnarray}
\begin{array}{ l l l l l l l l}
P  = \rho R T,  & \rho e_o = \frac{P}{\gamma -1} + \frac{1}{2} \rho u_i u_i,  &  \tau_{ij} =  \overbrace{\mu \cdot \left(\partial_{x_j} u_i + \partial_{x_i} u_j \right)}^{2S_{ij}} - \frac{2}{3} \mu \ \delta_{ij} \cdot \partial_{x_k} u_k, \\ \gamma = c_p/c_v, & R = c_p - c_v, &  q_i = \kappa \cdot\partial_{x_i} T
 \end{array}   \label{eq:}
\end{eqnarray}

\subsubsection{Initialization}

The vortex is initialized via velocity and temperature perturbations to a uniform background flow,
\begin{eqnarray}
\begin{array}{r c l}
\delta u  &=&   - \sqrt{R_{\infty}T_{\infty}} \left(\frac{\alpha}{2\pi}\right)(y - y_o) e^{\phi(1-r^2)} \\ \\
\delta v   &=&    \sqrt{R_{\infty}T_{\infty}} \left(\frac{\alpha}{2\pi}\right)(x - x_o) e^{\phi(1-r^2)} \\ \\
\delta T  &=&   - T_{\infty} \left[\frac{\alpha^2 (\gamma - 1)}{16\phi \gamma \pi^2}\right] e^{2\phi(1-r^2)}
\end{array} \label{eq:27}
\end{eqnarray}
where the radius, $r^2 = (x-x_o)^2 + (y-y_o)^2$, is defined about an origin $(x_o,y_o)$.
Here, the vortex is initialized on top of a background flow, ${\bf Q}_{\infty}$, which features velocity in the x-direction only (i.e., $v_{\infty} = w_{\infty} = 0$):
\begin{eqnarray}
\begin{array}{l l l }
\rho u_\infty = 200.0  \ \left[\frac{kg}{m^2 \cdot s}\right],  &
\rho_\infty = 1.0  \ \left[\frac{kg}{m^3}\right], &
\rho e_{o,\infty} = 305714.3 \ \left[\frac{kg}{m \cdot s^2}\right]
\end{array}
\end{eqnarray}
The isentropic relation, $P = P_{\infty}(T/T_{\infty})^{\gamma/(\gamma-1)}$, is then invoked along with the ideal gas law in order to complete the thermodynamic specifications.
A periodic domain is considered, and the parameters $\alpha = 4$ and $\phi = 1$ are chosen in Equation \ref{eq:27} in order to produce a $\sim 45 \%$ density perturbation and a vortex diameter of approximately $2.1 \ [m]$ (the vortex diameter being identified according to $\delta T /T_{\infty} = 0.001$).
The following results are calculated using the fourth-order DRP optimized stencil (CD04-7opt) for first derivatives, and employs the quadratic splitting of Feiereisen \emph{et al.} \cite{Feiereisen:1981} for each of the transport terms in density, momentum, and total energy:
\begin{eqnarray}
\partial_{x_j} \rho u_j \psi \approx  \frac{1}{2} \cdot  \delta_{x_j} \rho u_j \psi
+ \frac{1}{2} \cdot \left(\rho u_j \delta_{x_j} \psi + \psi \delta_{x_j} \rho u_j \right)  \ . \label{eq:29}
\end{eqnarray}
The convective splitting mitigates aliasing activity \cite{Kennedy:2008,Edoh:2020} and satisfies secondary conservation properties \cite{Coppola:2019}.
The classic fourth-order Runge-Kutta method is used for time integration. 
On all grids considered, the various filtering methods are tuned to a $4$ points-per-wave (PPW) resolution (i.e., $k_c\Delta x = 0.5\pi$) per the respective cut-off definitions of the Top-hat and sixth-order Tangent filter.

\subsubsection{Results}

\begin{figure} [h!]
 \centering
 \subfigure[]{ \label{fig:9a}
  \includegraphics[width=0.45\textwidth]{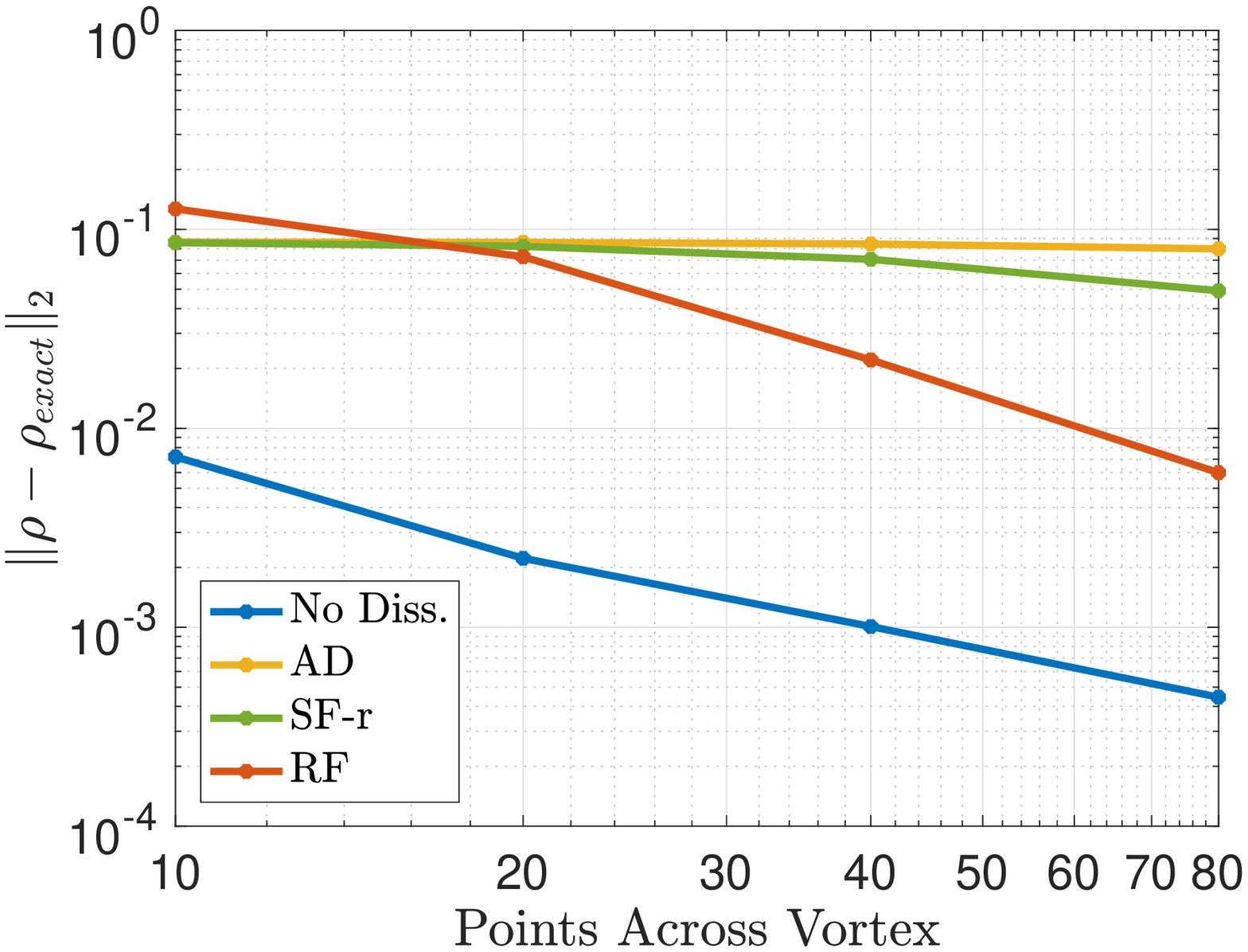}}
  \subfigure[]{ \label{fig:9b}
 \includegraphics[width=0.45\textwidth]{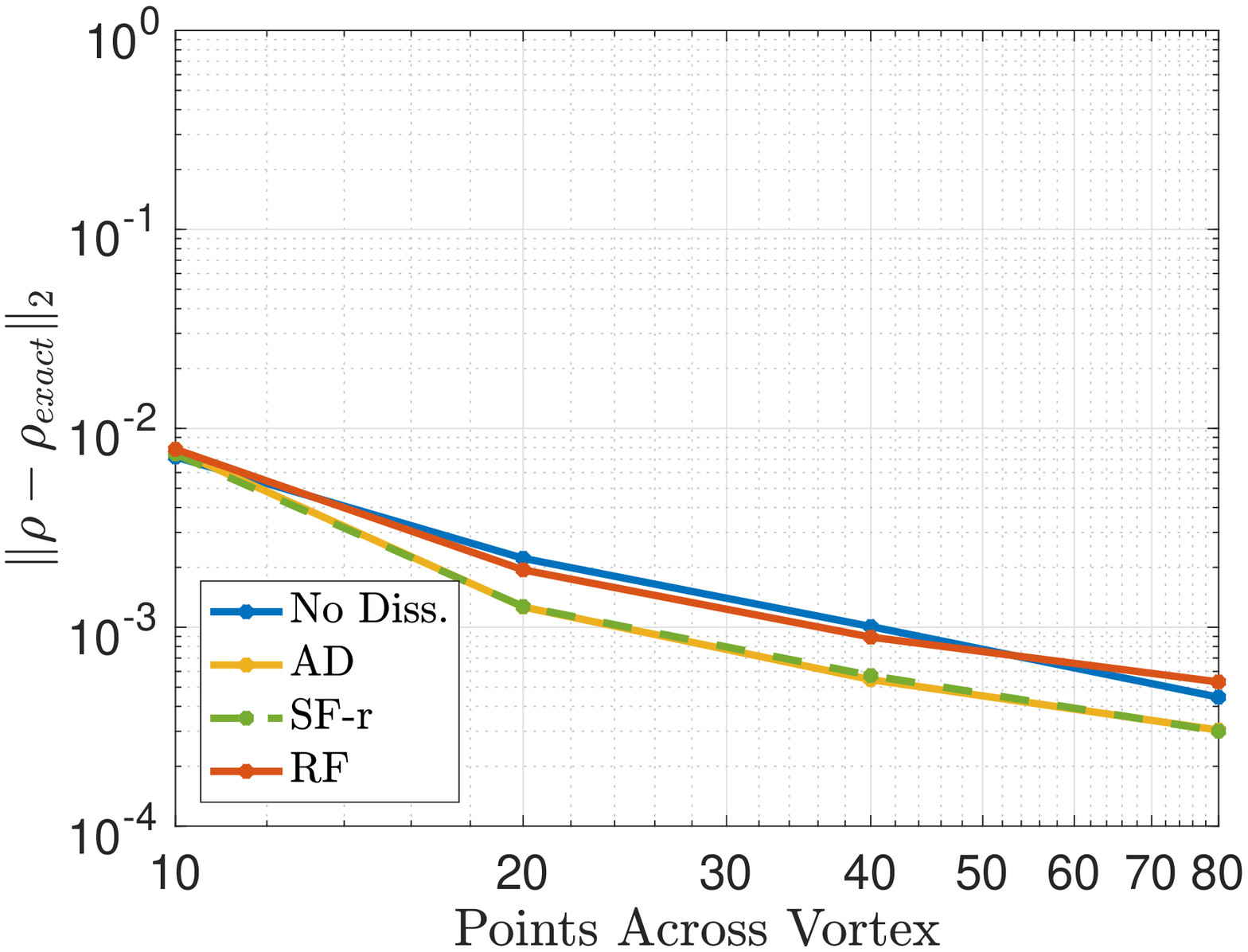}}
 \caption{Convergence of density solution after traveling two vortex widths and employing different filtering methods tuned to $(k_c\Delta x) = 0.5\pi$: a) Top-hat filter and b) sixth-order Tangent filter. }
 \label{fig:9}
\end{figure}

\begin{figure}[h!]
\centering
  \subfigure[] { \label{fig:10a}
 \begin{tabular}{c c c c c}
 No Fil. & RF &  SF-r \\
 \includegraphics[width=0.33\textwidth]{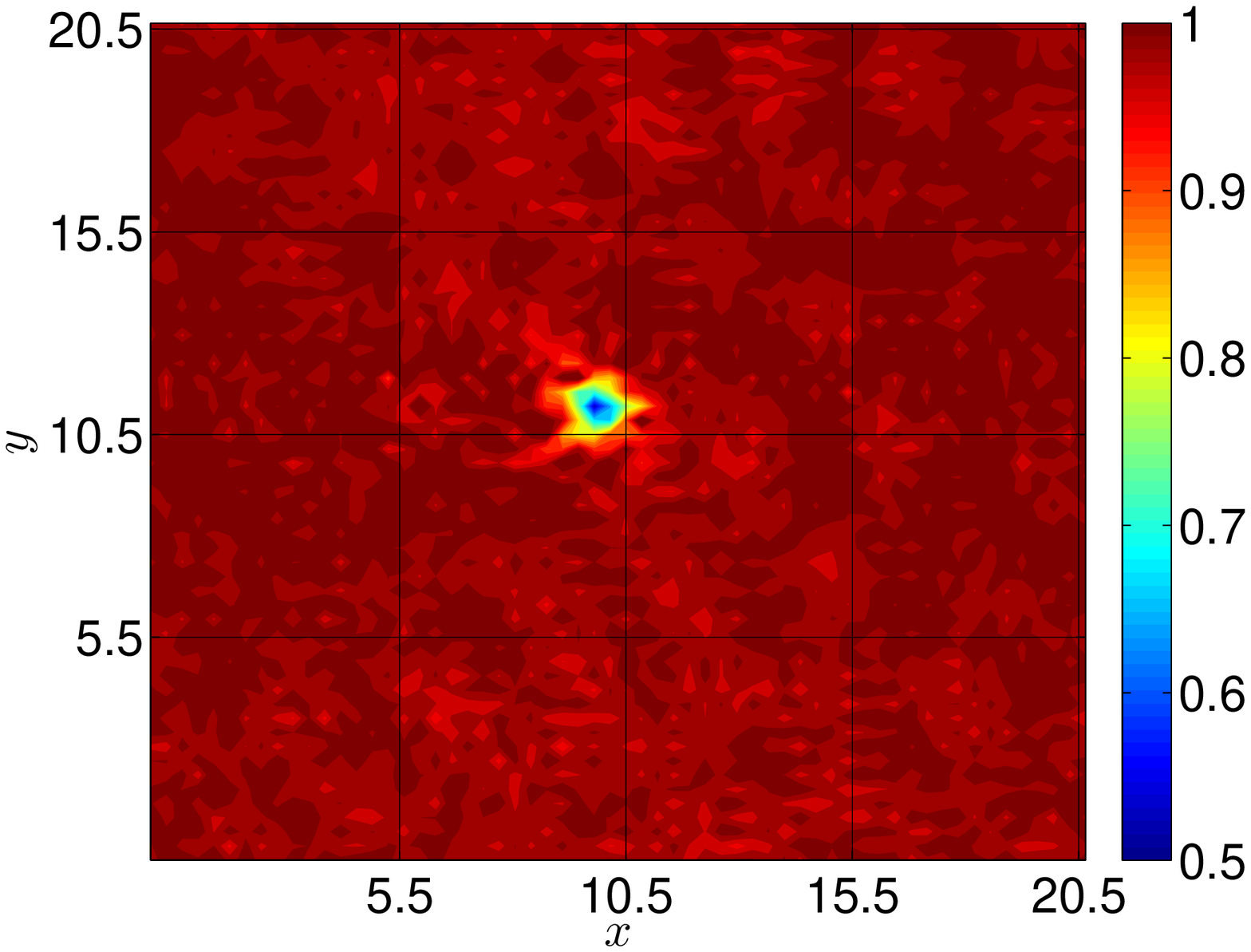} &
\includegraphics[width=0.33\textwidth]{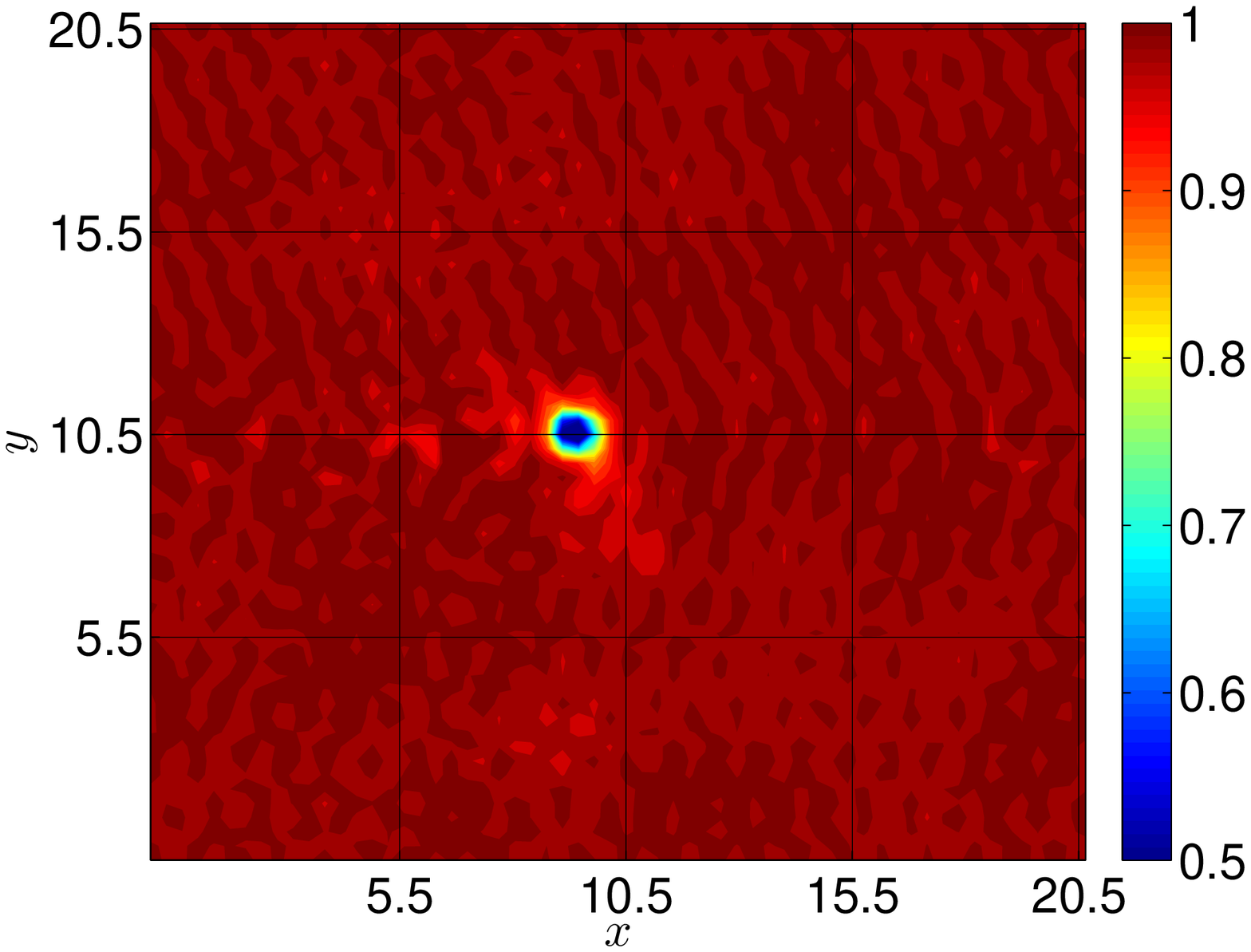}   &
\includegraphics[width=0.33\textwidth]{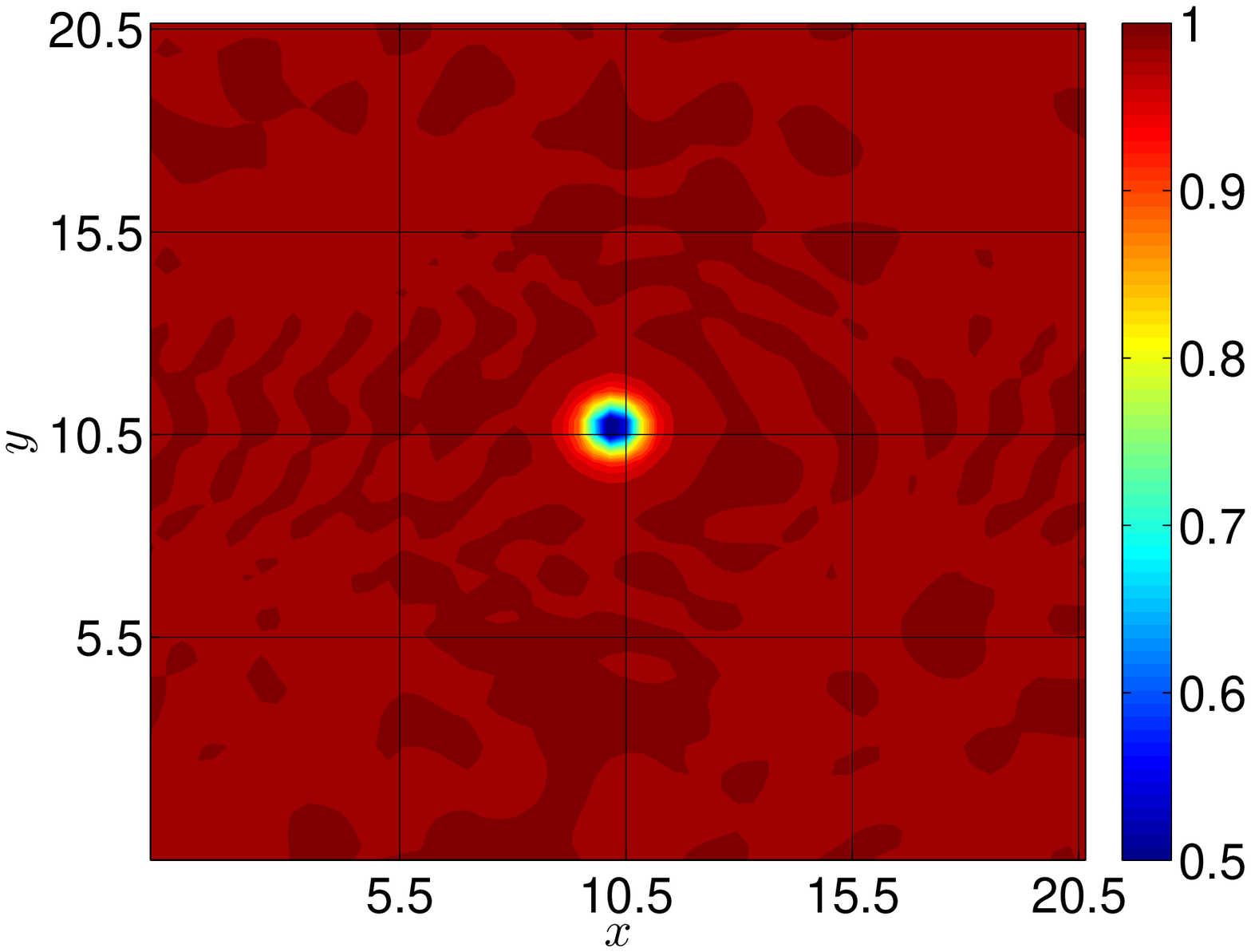}
 \end{tabular}}
 \subfigure[] { \label{fig:10b}
 \begin{tabular}{c c c c}
\includegraphics[width=0.33\textwidth]{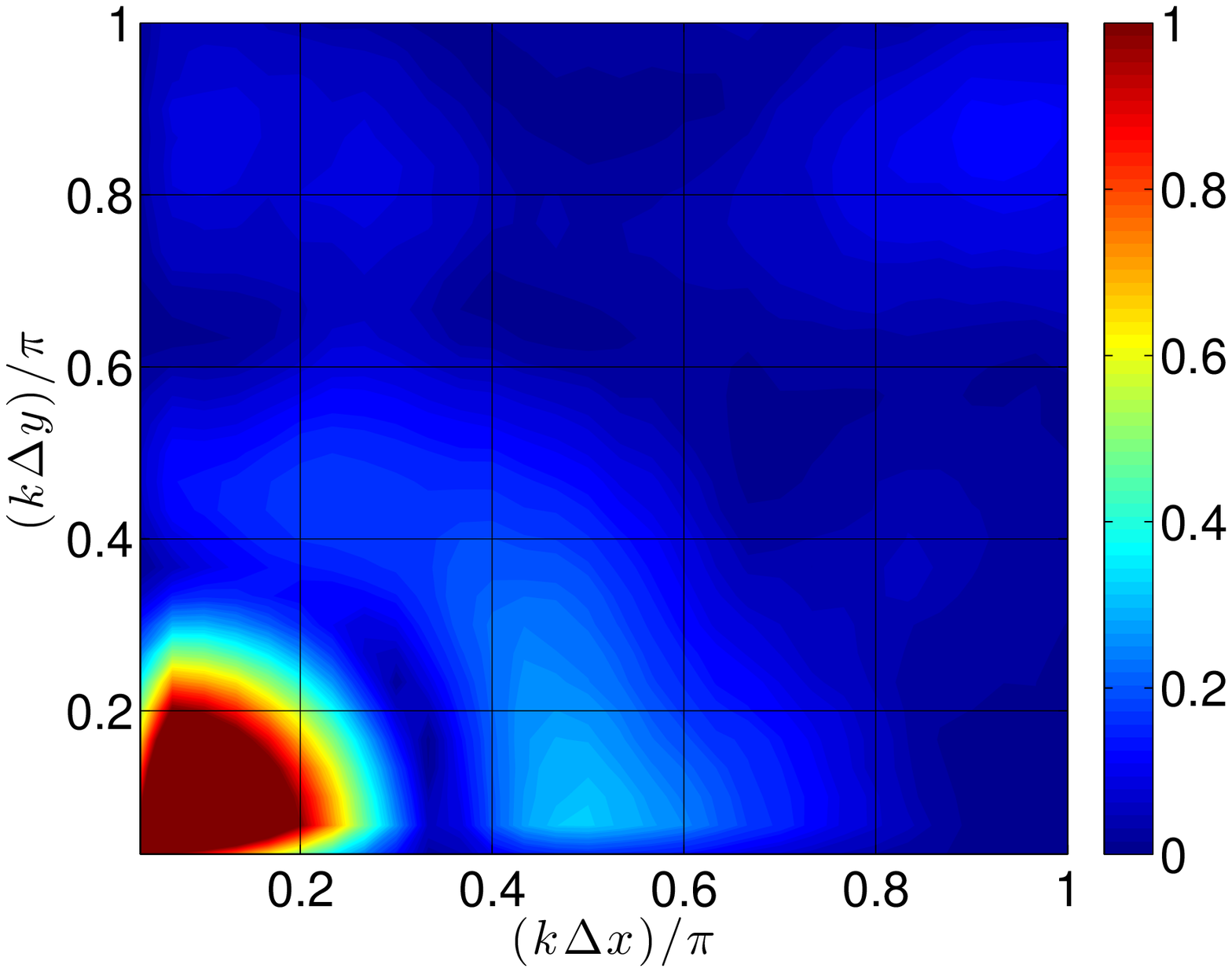} &
\includegraphics[width=0.33\textwidth]{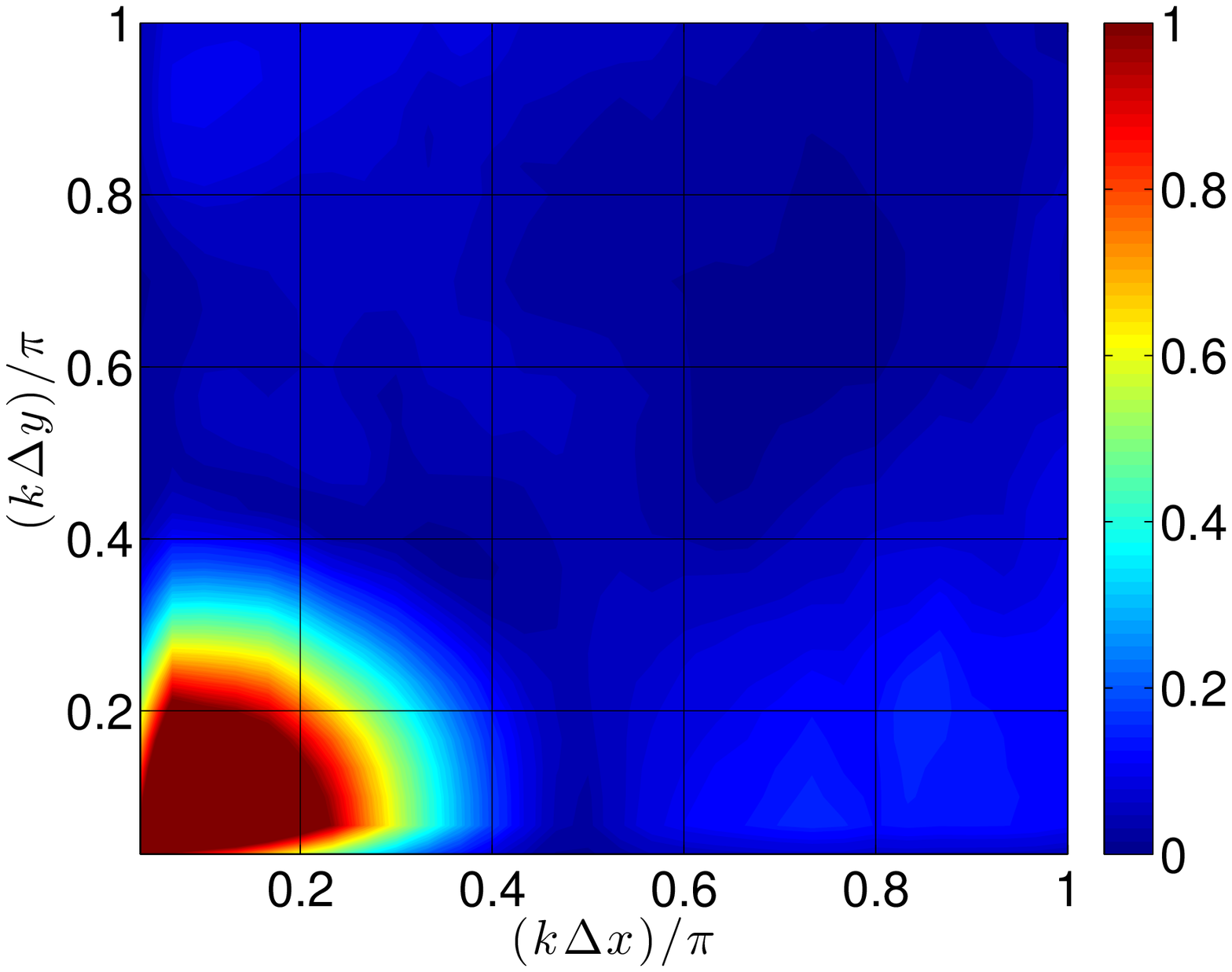} &
\includegraphics[width=0.33\textwidth]{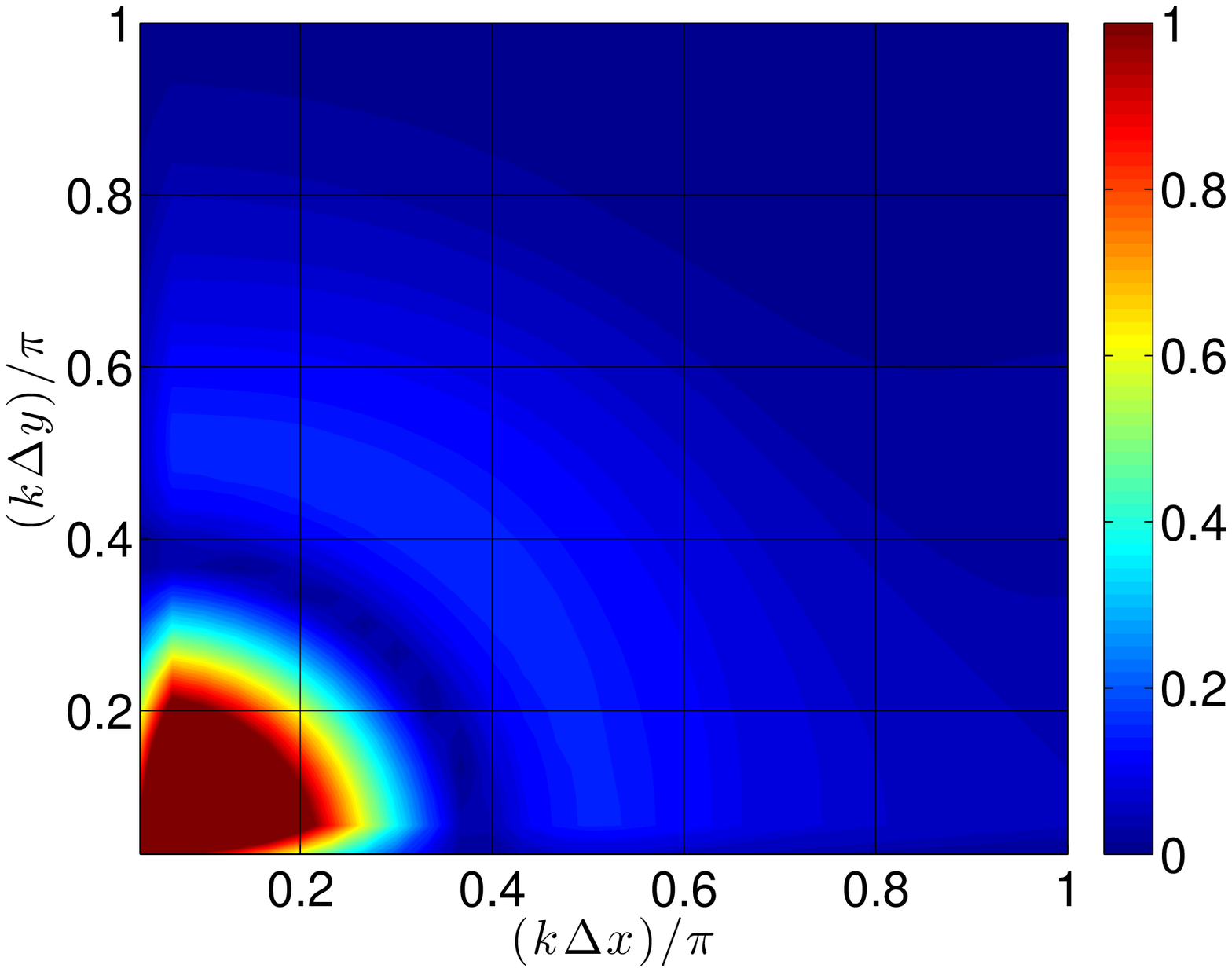}
 \end{tabular}}
 \caption{Solutions on the square $N = 60^2$ grid after traveling forty vortex widths and employing different filtering methods with sixth-order Tangent filter: a) density solution and b) normalized two-dimensional energy spectra of the velocity perturbations. }\label{fig:10}
 \end{figure}

Figure \ref{fig:9} presents convergence plots for the different filtering implementations as paired with the respective filters.
The data is calculated after two vortex widths traveled on a periodic domain of $L = 2.1 \ [m]$ (i.e., two vortex widths) with a time step of $\Delta t = 4.2 \times 10^{-6} \ [s]$ -- which corresponds to $CFL_{u_\infty,1D} \approx 0.016$ on the finest grid, therefore temporal effects are negligible at all resolutions considered and one can expect congruence between SF-r and AD results (as observed).
In the case of the Top-hat kernel (see Figure \ref{fig:9a},), the filter is overly dissipative relative to the spectral signature of the problem and therefore produces large errors that persist with the grid refinement. 
This issue is more pervasive with the SF-r/AD renditions of filtering than with RF. 
On the other hand, in Figure \ref{fig:9b}, the filtering utilizing the current Tangent stencil is shown to generally reduce numerical errors.
Here, the SF-r/AD implementations perform better than the RF implementation which admits more high wavenumber noise and phase error.

Figure \ref{fig:10} shows both the density solution and velocity spectra calculated with the Tangent filtering after forty vortex widths traveled on a $L = 21 \ [m]$ periodic square domain of $N = 60^2$ points (i.e., 10 points across the vortex width).
One notes that filtering generally mitigates the presence of small scales compared to the non-filtered solution.
However, it is noted that the RF solutions trails its expected location and has begun to accumulate high wavenumber content -- a consequence of its lack of a damping mechanism and its dispersive characteristics  with respect to the Euler system.
While the current observations are specific to the case and methods used, they corroborate the theoretical analysis from previous sections and highlight potential impact of the filter characteristics (e.g., scale-discriminant attenuation) relative to the different filtering methods.

\subsection{3D Navier-Stokes: Taylor-Green Vortex ($\text{Re} = 1600$)} \label{sec:3c}

The three-dimensional compressible Taylor-Green vortex (TGV) problem, governed by the 3D Navier-Stokes system (see Equation \ref{eq:25}), is now considered with the intent to assess the efficacy of the different filtering methods while refining the grid. 
The TGV is an unsteady vortex featuring the transfer of large-scale perturbations towards small-scale features, followed by an eventual decay.
The test case thus serves as more complex and relevant analogue to the previous viscous Burgers example.

Quantitative characterization of the flow is achieved by analyzing the volume-averaged kinetic energy $(E_k)$, its rate of change with respect to time $(\epsilon_{E_k})$, and the viscous dissipation $(\epsilon_{S_{ij}})$:
\begin{equation}
\begin{array}{c c c}
E_k = \frac{1}{\rho_0 \Omega} \int_{\Omega} \rho
\frac{u_i u_i}{2} d \Omega, &
\epsilon_{E_k} = -d_t E_k, &
\epsilon_{S_{ij}} = \frac{2\mu}{\rho_0 \Omega} \int_{\Omega} S_{ij} S_{ij} \ d\Omega
\end{array}
\end{equation}
Although $\epsilon_{E_k}$ characterizes the overall kinetic energy dissipataion, $\epsilon_{S_{ij}}$ represents the dissipation due to the viscosity terms and can also be used as a practical measure of the amount of small scale activity.
Upon sufficient spatial resolution (and assuming small compressibility effects such that $\partial_{x_i} u_i \approx 0$), one recovers the analytical result of $\epsilon_{E_k} \approx \epsilon_{S_{ij}}$, as suggested by the evolution equation for kinetic energy,
\begin{eqnarray}
\frac{1}{2} \partial_t (\rho u_i u_i) = u_i\partial_t (\rho u_i) - \frac{1}{2} u_i^2 \partial_t \rho
=  - \frac{1}{2} \partial_{x_j}  \rho u_j u_i u_i  - u_i \partial_{x_i} P + u_i  \partial_{x_j}\tau_{ij}  \ . \label{eq:33}
\end{eqnarray}
However, in the case of under-resolved grids, there can be a discrepancy between the two metrics which can be interpreted as the implied non-linear effects stemming from the numerical method \cite{Domaradzki:2015},
\begin{equation}
\epsilon_{\text{num}} = (\epsilon_{E_K} - \epsilon_{S_{ij}}) \ .
\end{equation}
This contribution characterizes the numerically-induced effects that linger in the discrete version of Equation \ref{eq:33}, as derived from the discretized primary equations.

\subsubsection{Impact of Filtering Implementations on the Kinetic Energy Dissipation Rate}

The explicit form of $\epsilon_{\text{num}}$ can be written for each of the filtering methods in order to better understand their respective behaviors.
In order to do this, we consider $\mathcal{R}_{\text{o,ex}}$ to be the original residual that is discretized by a high order method of maximal accuracy relative to the available grid resolution (i.e., an ``exact" reference).
Then $\mathcal{R}'$ is the modified residual associated with each filtering method.
The numerical errors of the primary equations therefore manifest in the kinetic energy equation as $\epsilon_{num}$, where
\begin{eqnarray}
\epsilon_{\text{num}} &=& \int_\Omega \left[ u_i \left(\mathcal{R}'^{(\rho u_i)} - \mathcal{R}_{\text{o,ex}}^{(\rho u_i)}  \right) - \frac{1}{2} u_i^2 \left(\mathcal{R}'^{(\rho)} - \mathcal{R}_{\text{o,ex}}^{(\rho)} \right) \right]d \Omega, \quad \text{with} \
 \left\{\begin{array}{r c l} \mathcal{R}'_{\text{AD}}   &=& \mathcal{R}_{\text{o}} +\mathcal{D}_{\text{fil}}(Q) \\ 
\mathcal{R}'_{\text{RF}} &=& \mathcal{G}_{\text{fil}}\mathcal{R}_{\text{o}} \end{array} \right.  \nonumber \\\end{eqnarray}
Note that in order to simplify the current exposition, the modified residual for solution filtering is taken as a limiting case of artificial dissipation (i.e., $\lim_{\Delta t \to 0} \mathcal{R}'_{SF} = \mathcal{R}'_{\text{AD}}$).
After defining the truncation error as
\begin{equation}
\mathcal{E}_{\text{trunc}} = \mathcal{R}_{\text{o}} - \mathcal{R}_{\text{o,ex}} \ ,
\end{equation}
then the numerical contributions to the kinetic energy dynamics for each of the filtering procedures are expressed as
\begin{eqnarray}
\epsilon_{\text{num,AD}}
&=& \int_\Omega \left[\left(u_i\mathcal{E}_{\text{trunc}}^{(\rho u_i)} - \frac{1}{2}u_i^2\mathcal{E}_{\text{trunc}}^{(\rho)}\right) + \left(u_i\mathcal{D}_{\text{fil}}\rho u_i - \frac{1}{2}u_i^2\mathcal{D}_{\text{fil}}\rho\right)  \right] d\Omega \label{eq:36} \\ \nonumber \\
\epsilon_{\text{num,RF}} &=& \int_\Omega \left[ \left(u_i\mathcal{E}_{\text{trunc}}^{(\rho u_i)} - \frac{1}{2}u_i^2\mathcal{E}_{\text{trunc}}^{(\rho)}\right) + \left(u_i\mathcal{D}_{\text{fil}}\mathcal{R}_{\text{o}}^{(\rho u_i)} - \frac{1}{2}u_i^2\mathcal{D}_{\text{fil}}\mathcal{R}_{\text{o}}^{(\rho)} \right)\right] d\Omega \label{eq:37} 
\end{eqnarray}
The discretization error, $\mathcal{E}_{\text{trunc}}$, is seen to induce an effect in the kinetic energy balance for each filtering method.
In the case of artificial dissipation (and thus, by mild extension, the re-scaled solution filtering), the filter-related terms serve primarily as a stabilizing sink term whose influence increases according to magnitude of the filter attenuation.
Meanwhile, the filter-related terms in $\epsilon_{\text{num}}$ for residual filtering are far more indefinite. 
Further substituting the identities $\mathcal{R}_{o} = (\mathcal{R}_{\text{o,ex}} + \mathcal{E}_{\text{trunc}})$ and $\mathcal{G}_{\text{fil}} = (I + \mathcal{D}_{\text{fil}})$ into Equation \ref{eq:37} yields
\begin{eqnarray}
\epsilon_{\text{num,RF}} &=& \int_\Omega \left[\left(u_i\mathcal{G}_{\text{fil}}\mathcal{E}_{\text{trunc}}^{(\rho u_i)} - \frac{1}{2}u_i^2\mathcal{G}_{\text{fil}}\mathcal{E}_{\text{trunc}}^{(\rho)}\right) + \left(u_i\mathcal{D}_{\text{fil}}\mathcal{R}_{\text{o,ex}}^{(\rho u_i)} - \frac{1}{2}u_i^2\mathcal{D}_{\text{fil}}\mathcal{R}_{\text{o,ex}}^{(\rho)}\right)\right] d\Omega \ ,  \label{eq:38}
\end{eqnarray}
which provides some additional insight with respect to residual filtering.
It suggests that a proper selection of the filter can recover $\mathcal{G}_{\text{fil}}\mathcal{E}_{\text{trunc}} \approx 0$ and eliminate the influence of numerical errors from the baseline method. 
To do this, the filter must be properly tuned (i.e., in terms of cut-off and spectral sharpness) relative to the resolution and aliasing properties (i.e., the numerical errors) of the discretization scheme.
The lingering errors are then tied to the dispersive effects of residual filtering and are primarily characterized by the filter choice and the amount of small-scale content.
The above observations highlight the nuanced non-linear differences associated with the respective filtering methods, here expressed with respect to the Navier-Stokes kinetic energy dynamics\footnote{Note that when using a kinetic energy preserving (KEP) formulation \cite{Kuya:2018,Coppola:2019,Edoh:2022,Coppola:2023} such as the current splitting of Equation \ref{eq:29}, then the first terms in Equations \ref{eq:36} and \ref{eq:37} are zero, $\int_\Omega \left[ \left(u_i\mathcal{E}_{\text{trunc}}^{(\rho u_i)} - \frac{1}{2}u_i^2\mathcal{E}_{\text{trunc}}^{(\rho)}\right) \right]d\Omega$.}.

\subsubsection{Initialization}

The initial conditions to the Taylor Green vortex are given as perturbations to a base flow on a triply periodic domain, $-\pi L \le x,y,x \le \pi L$:
\begin{eqnarray}
\begin{array} {l l l}
\delta u &=& V_0 \sin\left(\frac{x}{L}\right) \cos\left(\frac{y}{L}\right) \cos\left(\frac{z}{L}\right) \\ \\
\delta v &=& -V_0 \cos\left(\frac{x}{L}\right) \sin\left(\frac{y}{L}\right) \cos\left(\frac{z}{L}\right) \\ \\
\delta P &=&  \frac{\rho_\infty V_0^2}{16} \left[ \cos\left(\frac{2x}{L}\right) + \cos\left(\frac{2y}{L}\right) \right]\left[ \cos\left(\frac{2z}{L}\right) + 2\right]
\end{array}
\end{eqnarray}
The problem is further prescribed by the following non-dimensional parameters:
\begin{eqnarray}
\begin{array}{c c c c}
{Re} = (\rho_\infty V_0 L)/\mu = 1600, &
{M}_0 = (V_0/c_0) = 0.1,  &
{Pr} = (\mu c_p)/\kappa = 0.71, &
\gamma = c_p/c_v = 1.4
\end{array}
  \label{eq:}
\end{eqnarray}

The following calculations employ a second-order stencil for the viscous terms and the optimized scheme CD04-7opt for the convective terms, which are quadratically split per Equation \ref{eq:29}.
Time integration is done via the classic fourth-order Runge-Kutta method with a time step size of $\Delta t= t_c/10^4$, where $t_c = L/V_o$ is the characteristic time scale.
Relative to the reference grid ($N = 512^3$), this yields $CFL_{V_o,1D} \approx 0.01$. 
Therefore temporal effects are minimal and differences between the AD and SF-r approaches are negligible, and the AD results are not reported henceforth.
Note that no sub-filter closure modeling is incorporated herein and that the current emphasis is on the behavior of the filtering implementations with regards to enforcing a target filter width and reducing numerical effects.

In the following, an effective resolution of $N = 64^3$ is applied with the Top-hat and Tangent filters per their respective cut-off definitions.
The Top-hat filter, however, is overly dissipative for the current resolutions and filter width when applied with SF-r/AD.
This hinders the initial perturbations (i.e., the flow quickly``laminarizes" and relaxes to the background conditions).
Therefore we omit the results based on the Top-hat filter employed with SF-r/AD.
Furthermore, the current RF implementation omits the diffusion terms $\partial_{x_j}{\bf E}^{(v)}_j$ as part of the residual filtering in the case of both the Top-hat and Tangent filters.
As previously explained and demonstrated, this modification is necessary for the numerical stability of the RF Top-hat method; and the RF Tangent method is implemented analogously for congruity in the results.

\subsubsection{Results}

Figures \ref{fig:11} and \ref{fig:12} show volumetric plots of vorticity for the RF and SF-r calculations for the Top-hat and Tangent filters, respectively.
These are compared to a reference solution that is filtered at the given instant in time.
One notices that RF features smaller structures than the provided filtered reference while SF yields a coarser effective resolution.
This suggests, once again, that RF tends to under-compensate the anticipated filter width while SF-r/AD tend to over-compensate for it.
Figure \ref{fig:13} further confirms these trends relative to the velocity spectra of the solutions.
The lack of an active attenuation mechanism in RF is seen to admit significant content past the anticipated cut-off -- more so for the Top-hat filter compared to the scale-discriminant Tangent filter.
Meanwhile the SF-r computation is shown to be overly dissipative compared to the target.

 \begin{figure}[h!]
 \centering
      \subfigure[Inst. filtered ref., Top-hat]{\label{fig:11a}
 \includegraphics[width=0.3\textwidth]{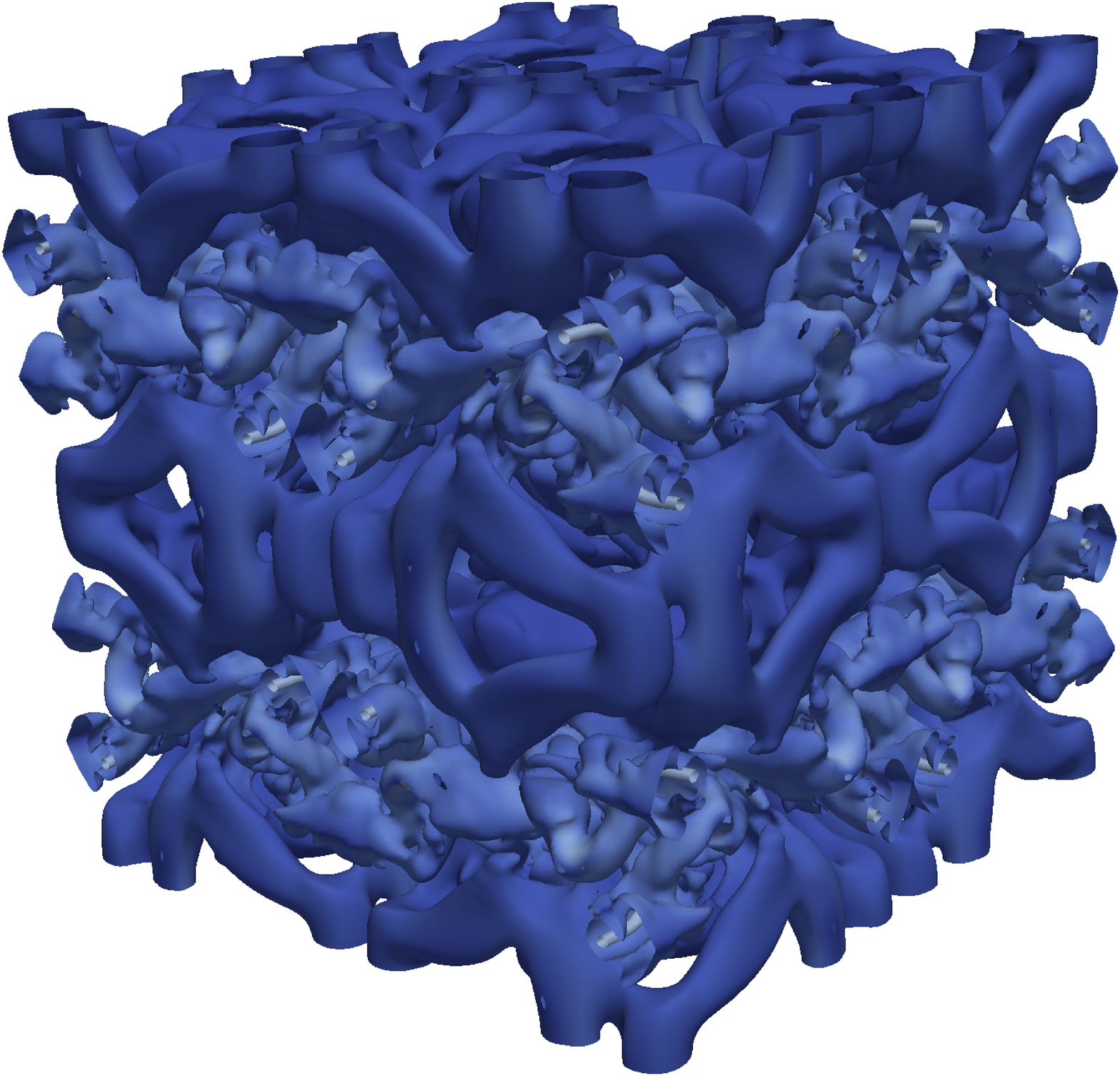}}
    \subfigure[Top-hat + RF]{\label{fig:11b}
 \includegraphics[width=0.3\textwidth]{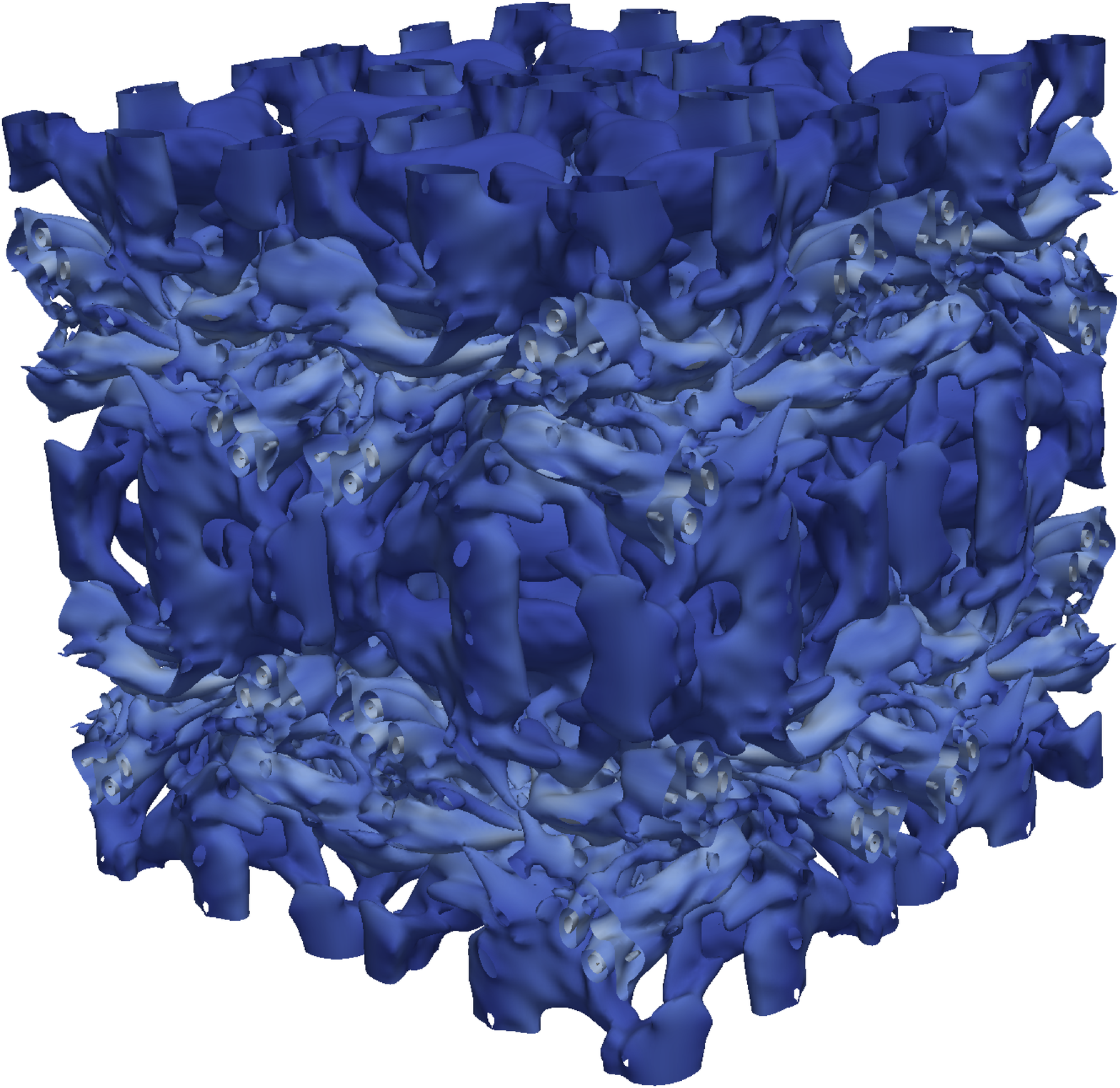}}
 \caption{Iso-surface of positive Q-criterion \cite{Haller:2005} (colored by vorticity magnitude) during time of vortical breakdown ($t = 9t_c$) for $N=192^3$ solutions with the Top-hat filter tuned to a $N=64^3$ resolution: a) instantaneously filtered reference and b) residual filtering. Note: Top-hat solution filtering is not shown because it is overly dissipative and quickly laminarizes.} \label{fig:11}
  \end{figure}
 \begin{figure}[h!]
 \centering
    \subfigure[Inst. filtered ref., Tan06]{\label{fig:12a}
 \includegraphics[width=0.3\textwidth]{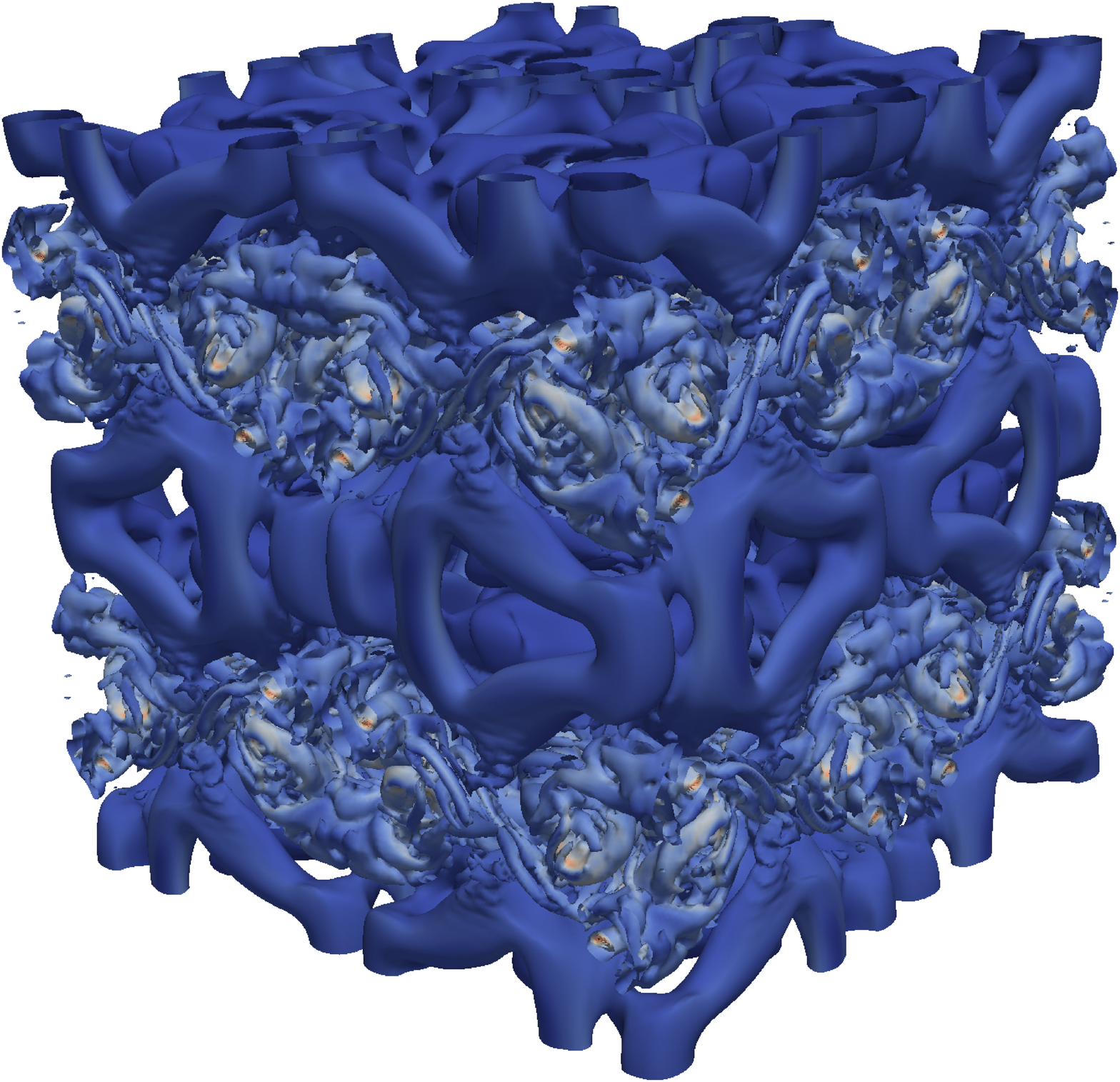}}
   \subfigure[Tan06 + RF]{\label{fig:12b}
 \includegraphics[width=0.3\textwidth]{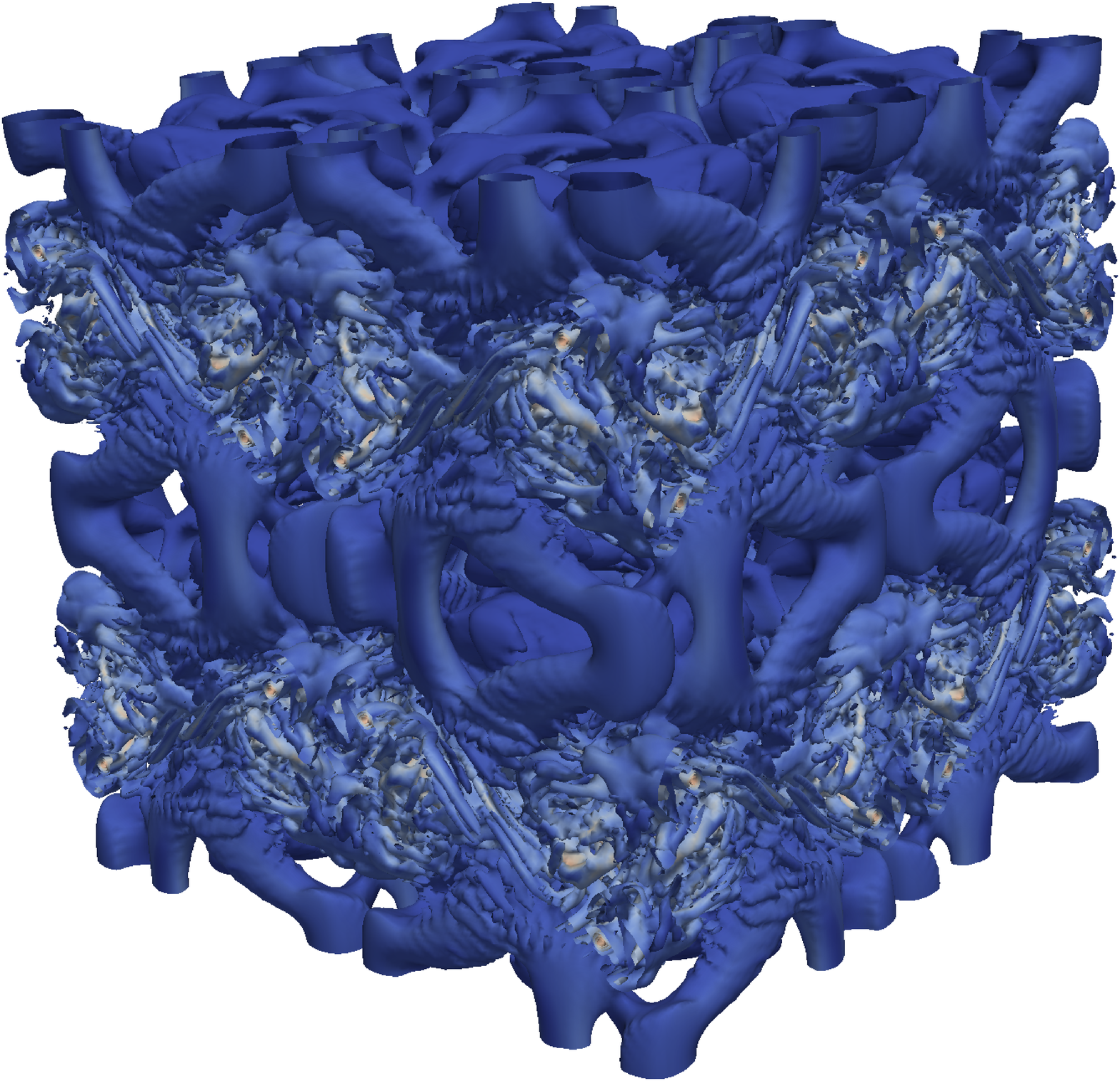}}
    \subfigure[Tan06 + SF-r]{\label{fig:12c}
 \includegraphics[width=0.3\textwidth]{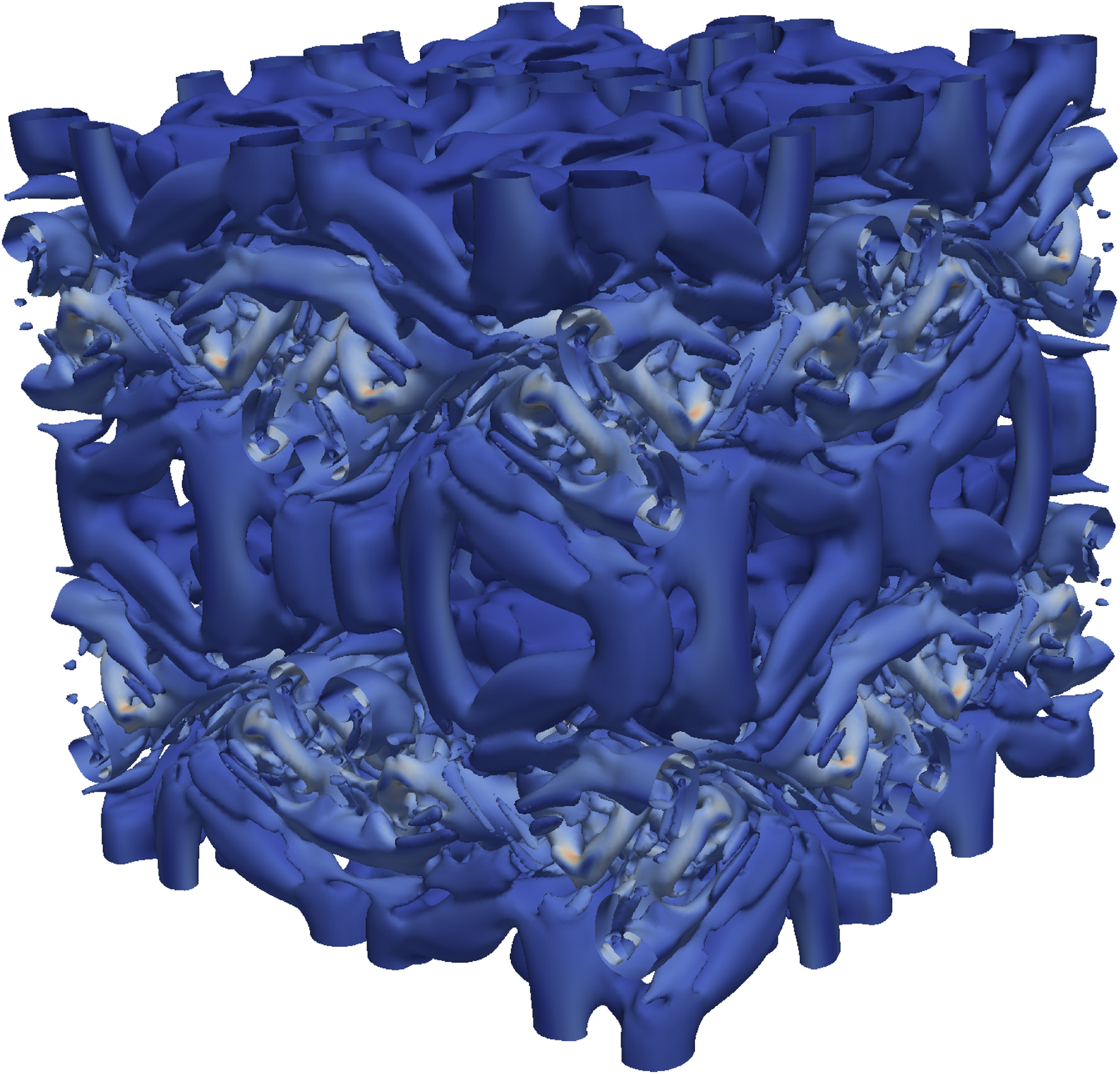}}
 \caption{Iso-surface of positive Q-criterion \cite{Haller:2005} (colored by vorticity magnitude) during time of vortical breakdown ($t = 9t_c$) for $192^3$ solutions with the sixth-order Tangent filter tuned to a $64^3$ resolution: a) instantaneously filtered reference, b) residual filtering, and c) temporally-consistent solution filtering.} \label{fig:12}
 \end{figure}
  \begin{figure}[h!]
  \centering
  \subfigure[Top-hat]{\label{fig:13a}
 \includegraphics[width=0.45\textwidth]{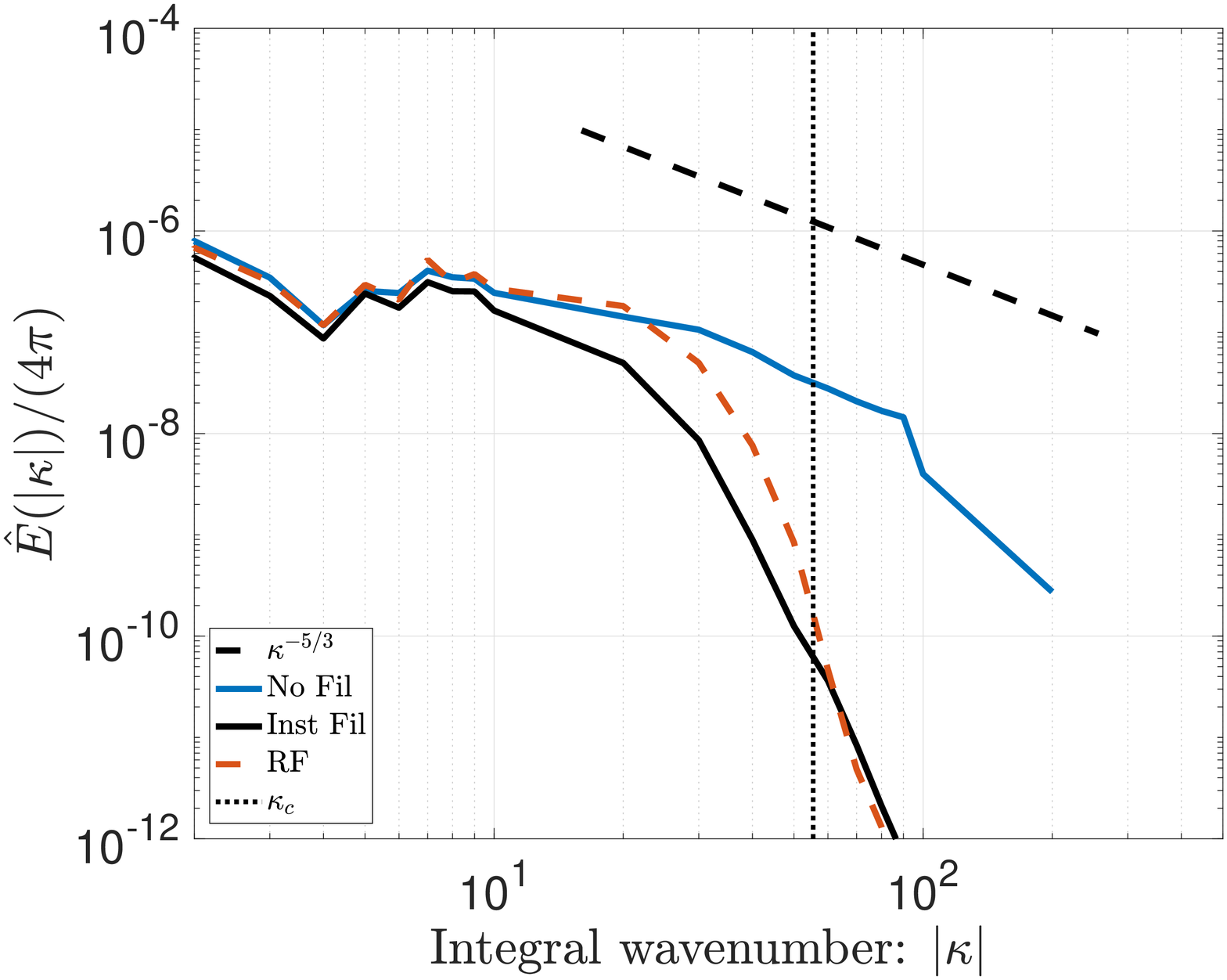}}
 \subfigure[Tan06]{\label{fig:13b}
 \includegraphics[width=0.45\textwidth]{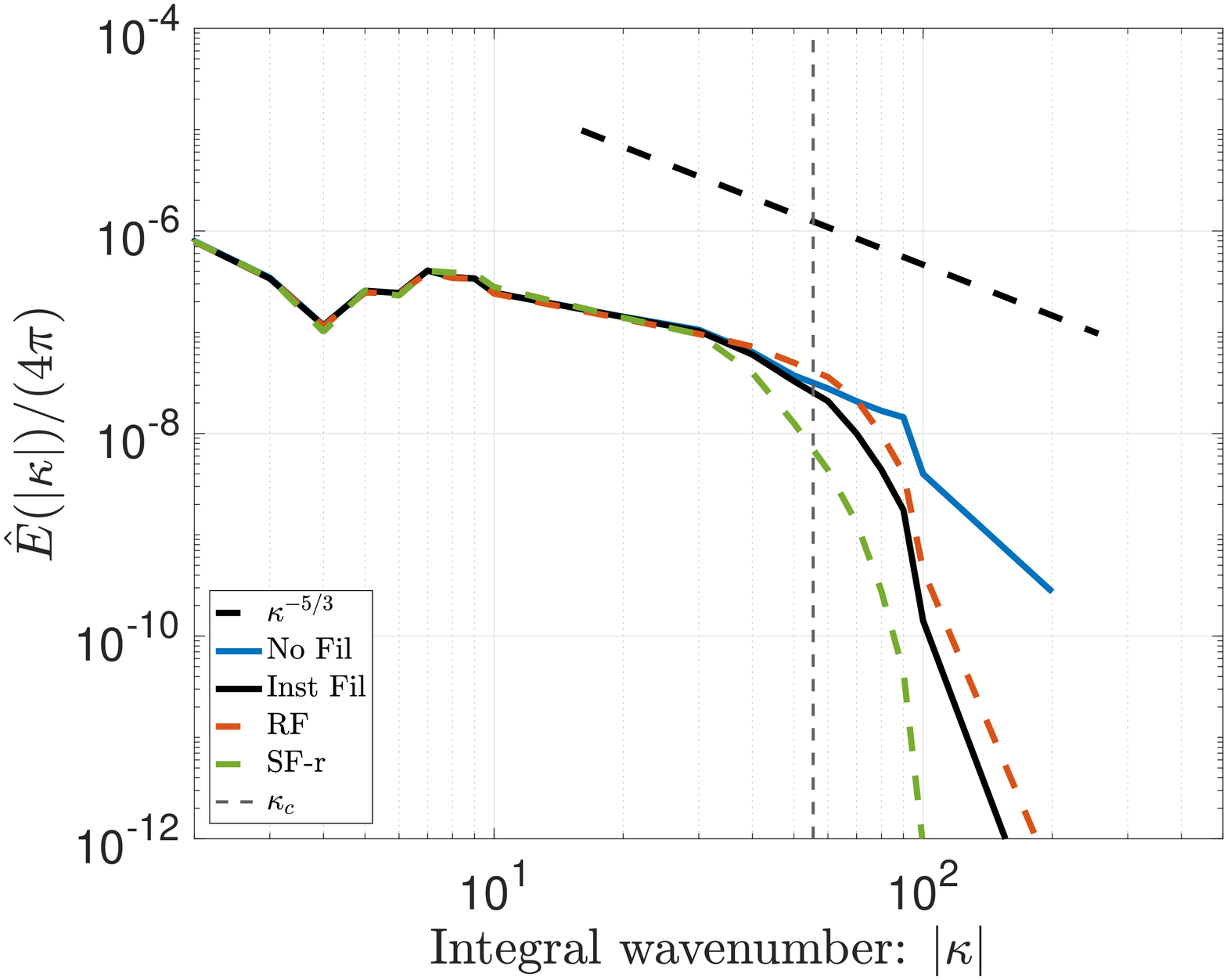}}
 \caption{Turbulent kinetic energy spectrum of $N=192^3$ solutions during vortical breakdown ($t = 9t_c$), comparing un-filtered and instantaneously filtered references to filtering methods as tuned to an effective resolution of $N=64^3$ corresponding to cut-off wavenumber $\kappa_c$: a) Top-hat filter and b) sixth-order Tangent filter.} \label{fig:13} 
 \end{figure}

 \begin{figure}[h!]
 \centering
 \subfigure[]{\label{fig:14a}
 \includegraphics[width=0.32\textwidth]{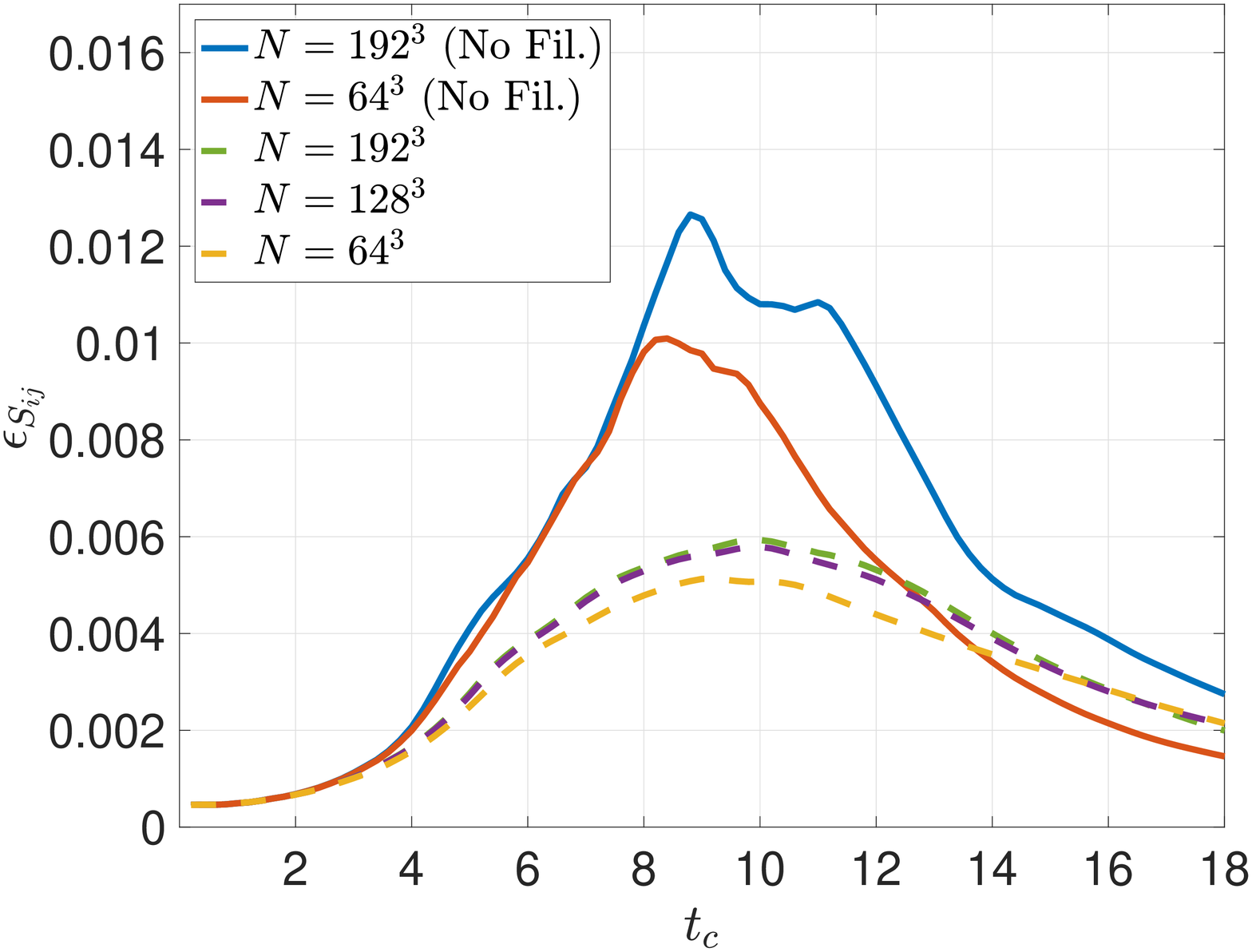}}
 \subfigure[]{\label{fig:14b}
 \includegraphics[width=0.32\textwidth]{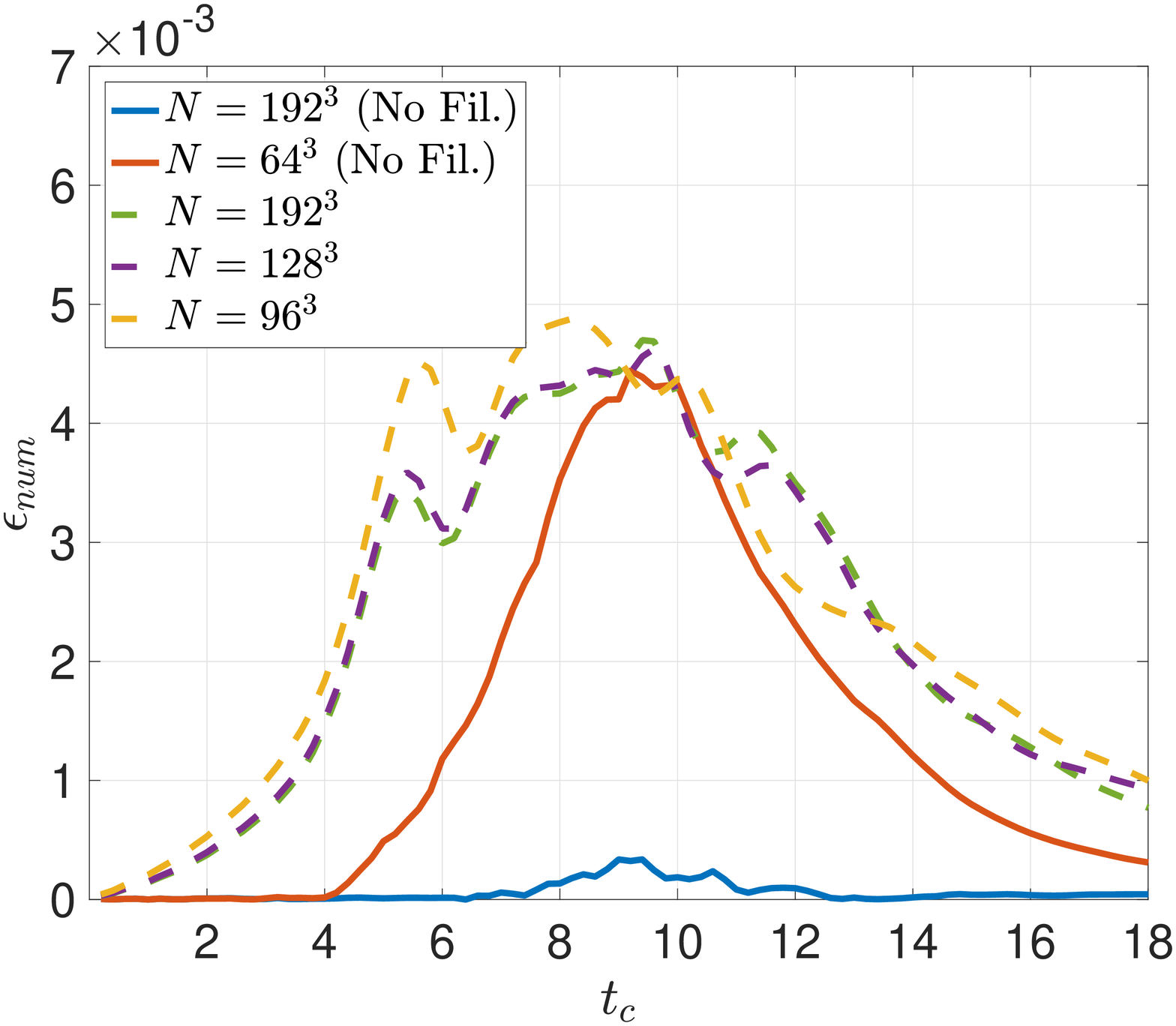}}
  \subfigure[]{\label{fig:14c}
 \includegraphics[width=0.32\textwidth]{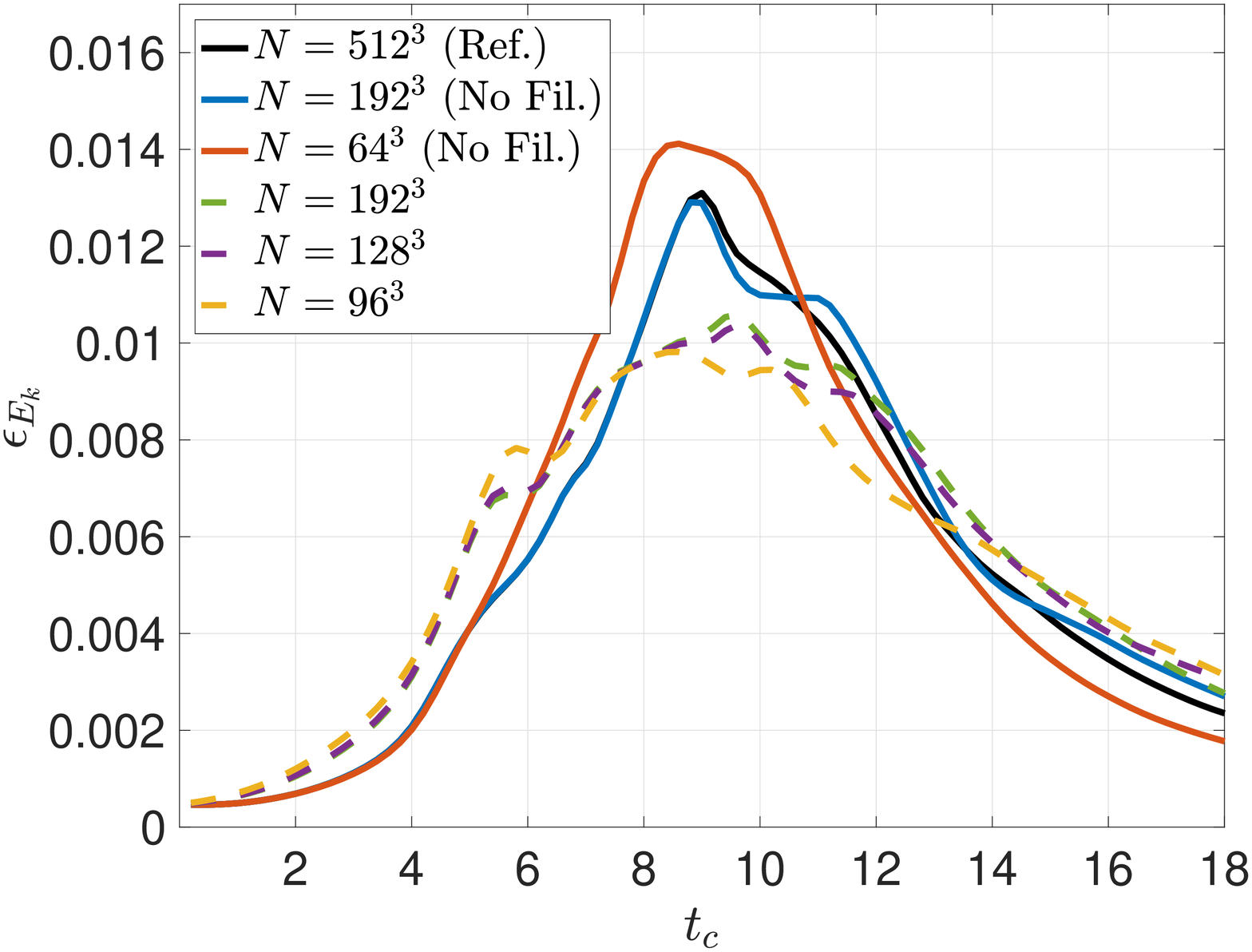}}
 \caption{Kinetic energy dissipation rates for the Top-hat filter with residual filtering (RF) tuned to a $N = 64^3$ cut-off: a) $\epsilon_{S_{ij}}$, b) $|\epsilon_{\text{num}}| = |\epsilon_{E_K} - \epsilon_{S_{ij}}|$ and c) $\epsilon_{E_k} = -d_t E_k$. } \label{fig:14}
\vspace{5mm}
 \centering
 \centering
  \subfigure[]{\label{fig:15a}
 \includegraphics[width=0.32\textwidth]{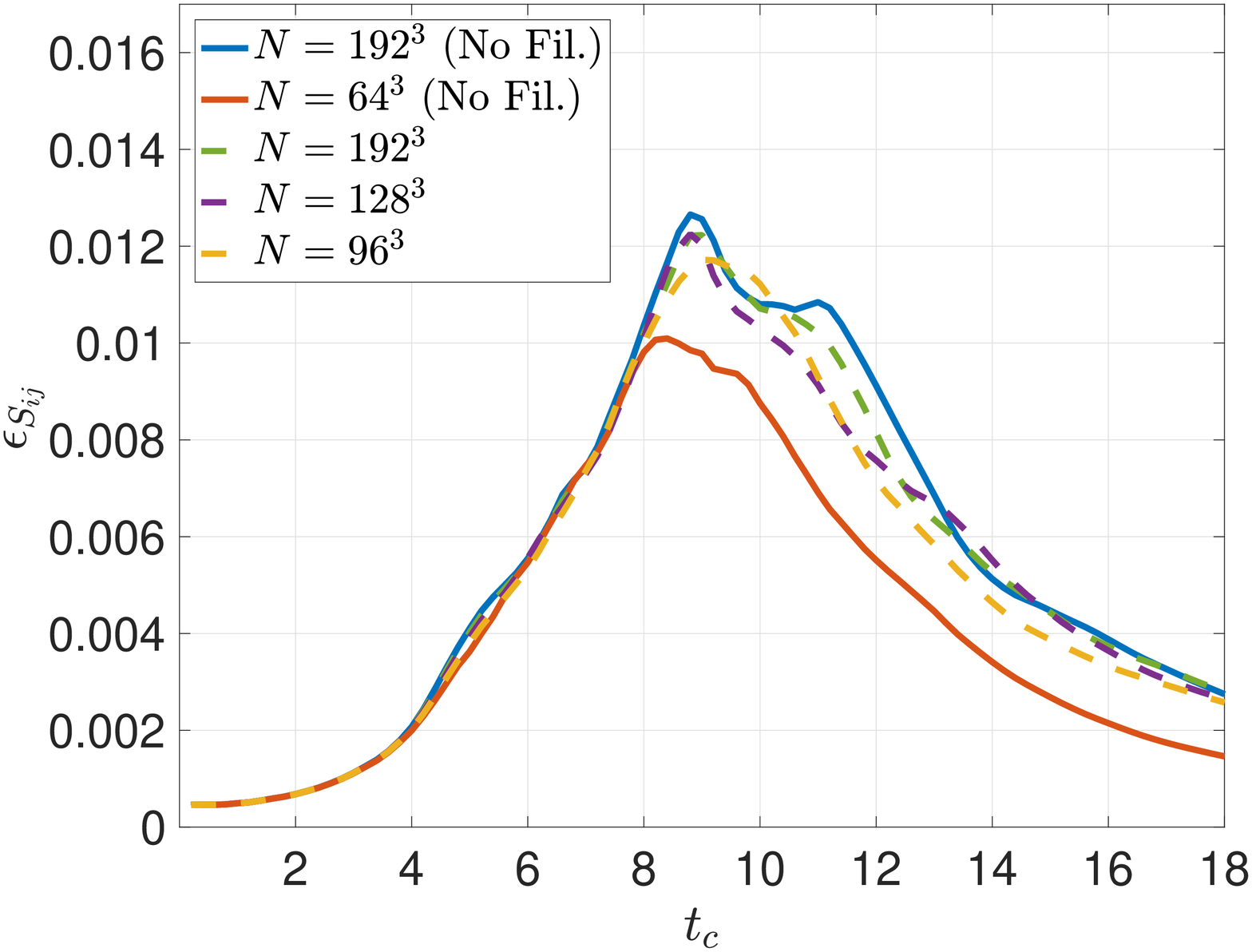}}
  \subfigure[]{\label{fig:15b}
 \includegraphics[width=0.32\textwidth]{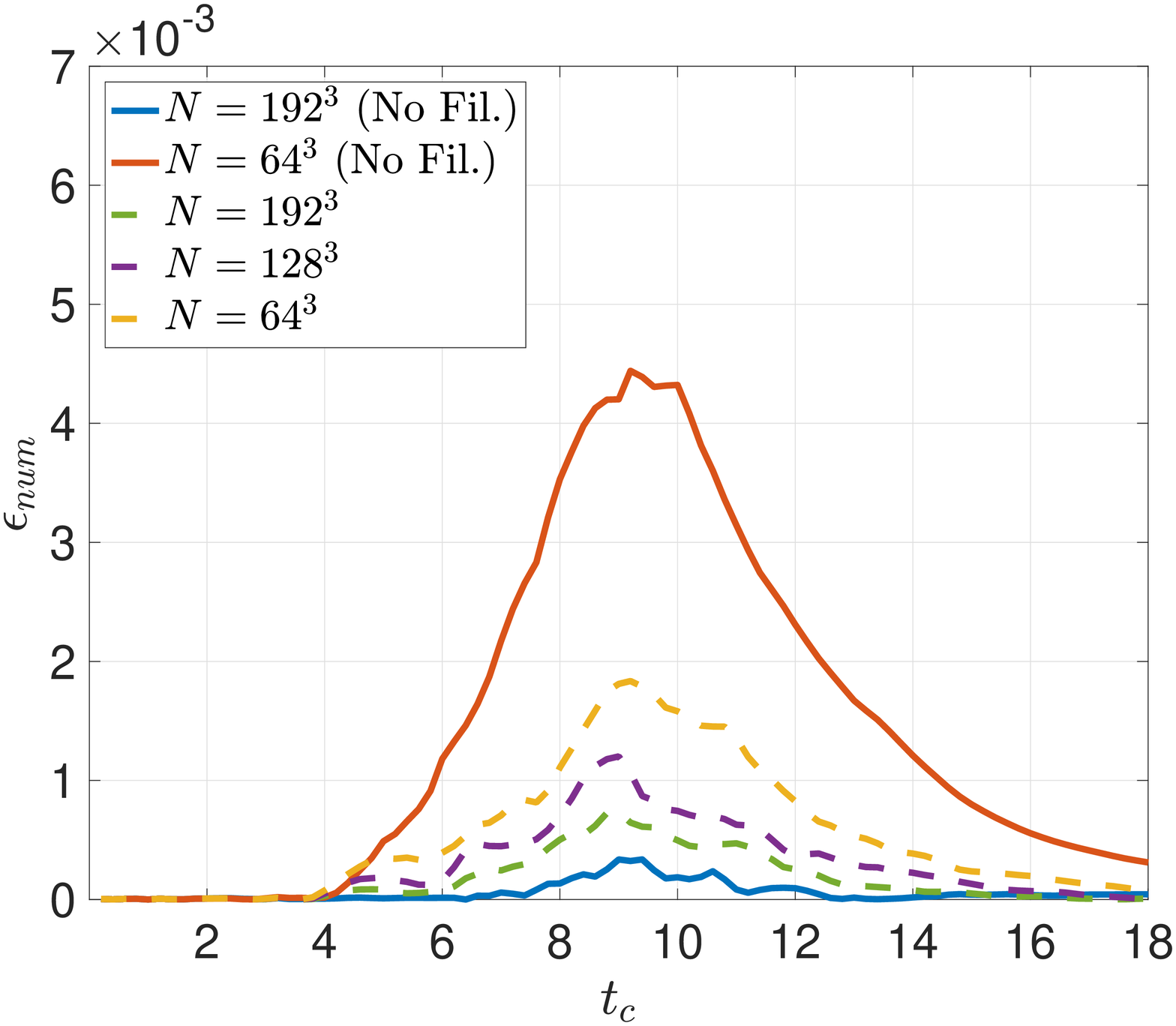}}
  \subfigure[]{\label{fig:15c}
 \includegraphics[width=0.32\textwidth]{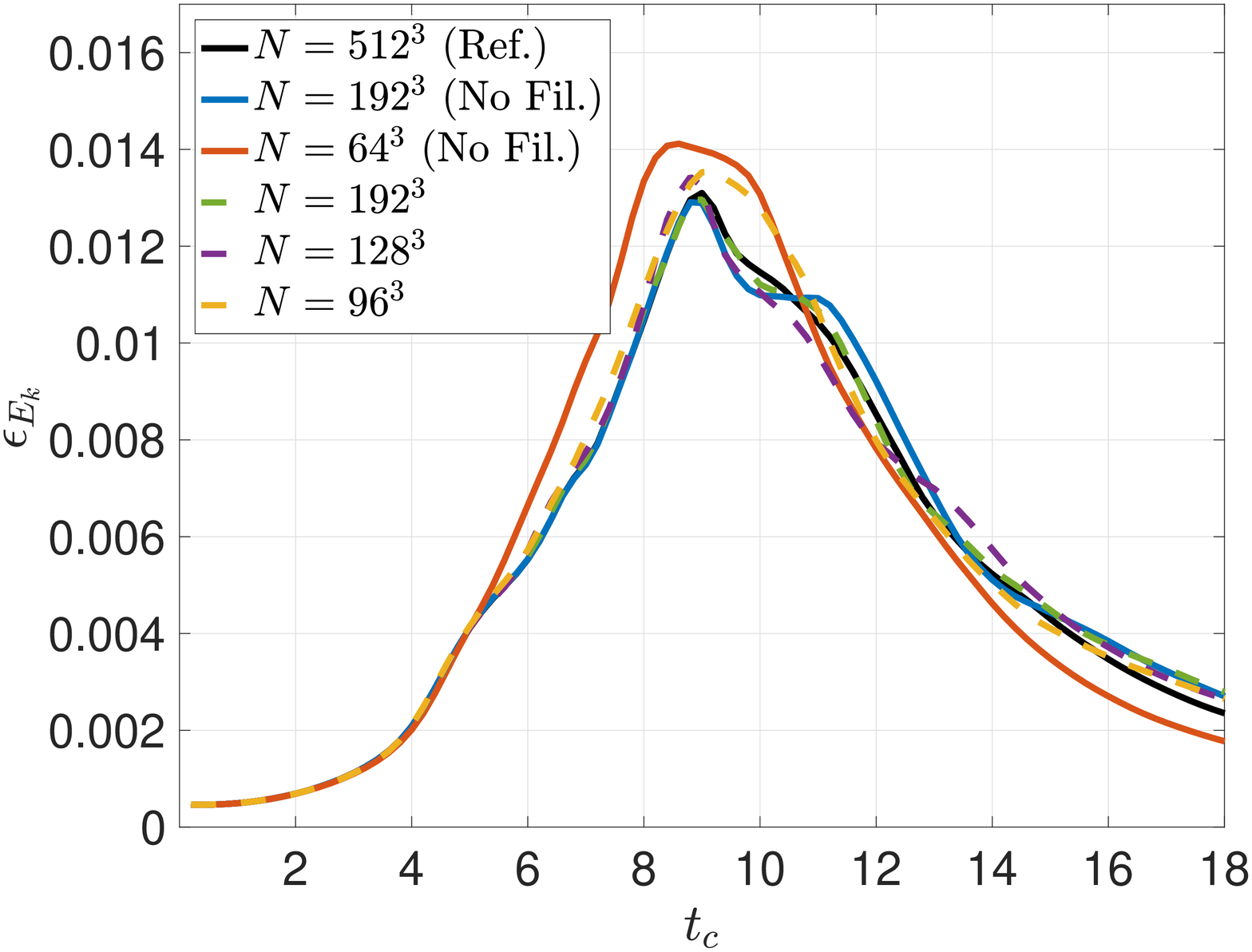}}
 \caption{Kinetic energy dissipation rates for the sixth-order Tangent filter with residual filtering (RF) tuned to a $N = 64^3$ cut-off: a) $\epsilon_{S_{ij}}$, b) $|\epsilon_{\text{num}}| = |\epsilon_{E_K} - \epsilon_{S_{ij}}|$ and c) $\epsilon_{E_k} = -d_t E_k$. } \label{fig:15}
\vspace{5mm}
  \subfigure[]{\label{fig:16a}
 \includegraphics[width=0.32\textwidth]{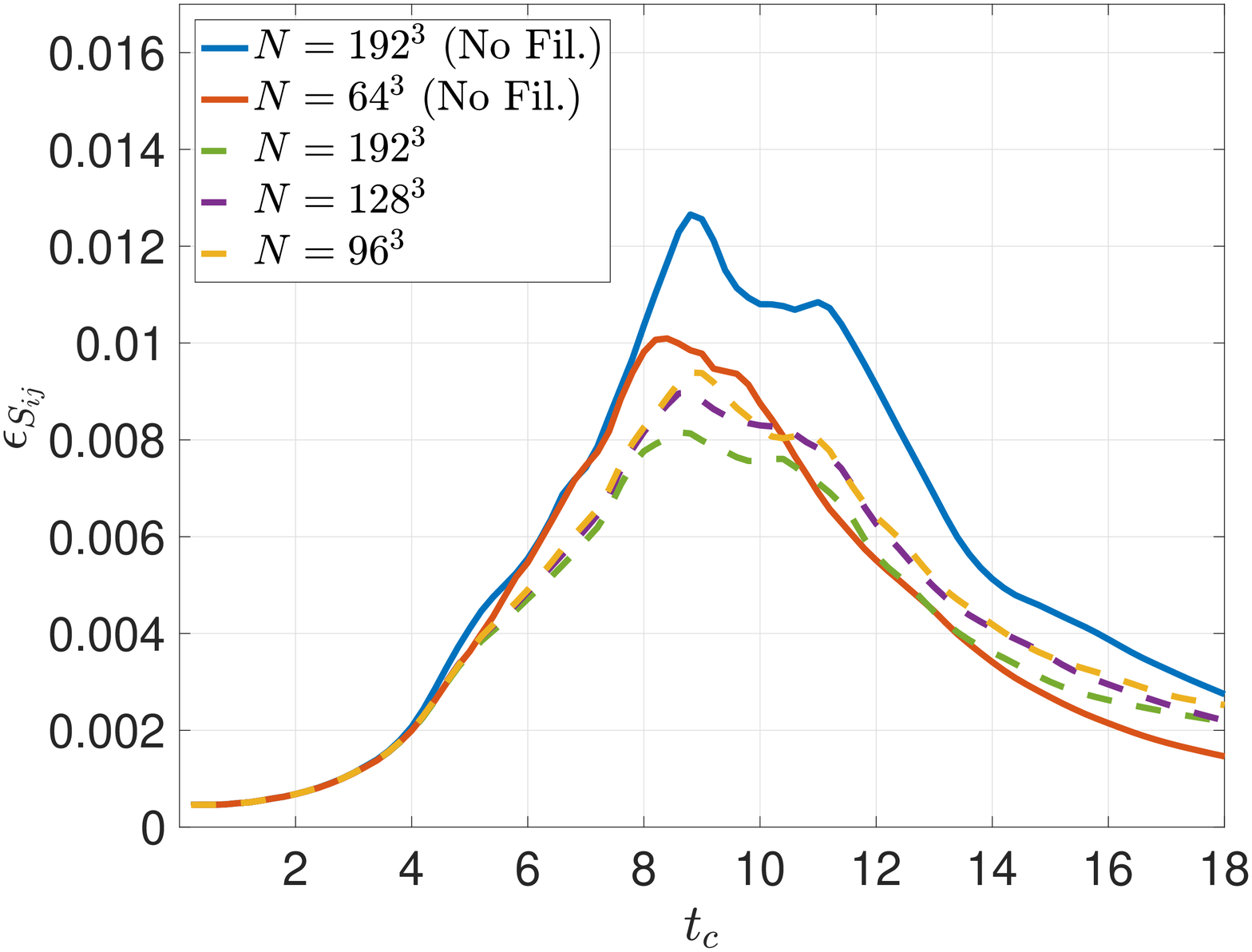}}
  \subfigure[]{\label{fig:16b}
 \includegraphics[width=0.32\textwidth]{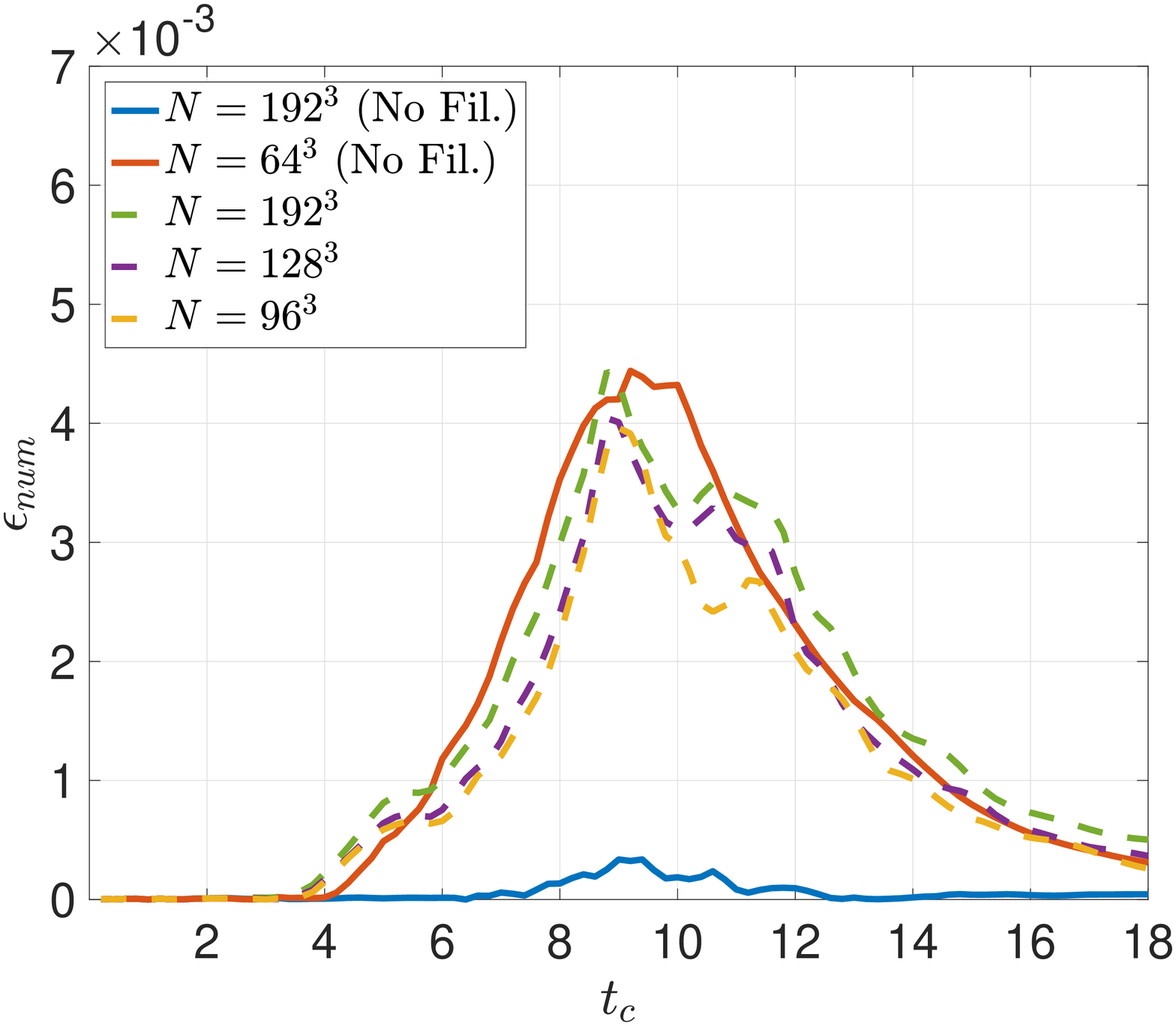}}
   \subfigure[]{\label{fig:16c}
 \includegraphics[width=0.32\textwidth]{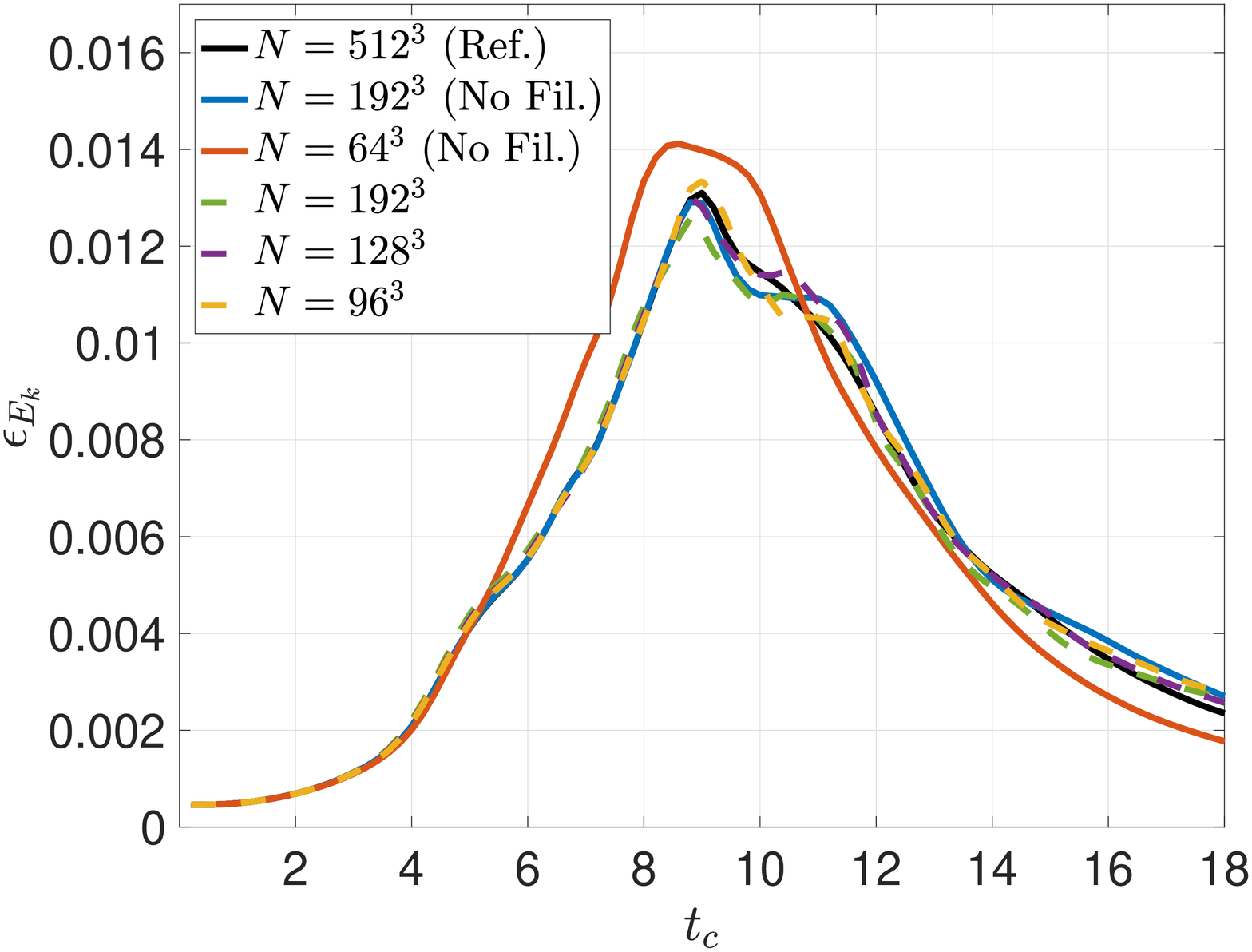}}
 \caption{Kinetic energy dissipation rates for the sixth-order Tangent filter with temporally-consistent re-scaled solution filtering (SF-r) tuned to a $N = 64^3$ cut-off: a) $\epsilon_{S_{ij}}$, b) $|\epsilon_{\text{num}}| = |\epsilon_{E_K} - \epsilon_{S_{ij}}|$ and c) $\epsilon_{E_k} = -d_t E_k$.} \label{fig:16}
\end{figure}

Figure \ref{fig:14} presents the different kinetic energy dissipation metrics for the Top-hat filter applied via RF with an effective cut-off resolution of $N = 64^3$.
The strain-based metric $\epsilon_{S_{ij}}$ in Figure \ref{fig:14a}, which is indicative of small-scale activity, confirms that the filtering hinders the generation of high wavenumber content.
The RF method performs consistently upon refinement of the mesh whilst maintaining the same physical filter width.
Inspecting the contribution of numerical effects (e.g., discretization error, filter-induced contributions) to the dissipation rate via the $\epsilon_{\text{num}}$ metric in Figure \ref{fig:14b} communicates that the solution dynamics are dominated by the low-order Top-hat response, even upon mesh refinement.
In other words, $\epsilon_{num,RF}$ in Equation \ref{eq:38} remains substantial due to the form of the Top-hat filter attenuation, $\mathcal{D}_{\text{fil}}$, which is very broad in spectral space and not scale-discriminant.
These characteristics should be taken into account when pairing LES closure models with such a formulation.
Overall, the Top-hat RF method strongly mollifies the original dynamics due to the filter response having an effect over a large range of scales, as anticipated from the theoretical analysis.
This yields a slow down of the cascading process.
Figure \ref{fig:14c} plots the total kinetic energy dissipation rate $\epsilon_{E_k}$ and shows a mild delay in the peak as well as a lower amplitude resulting from the slower flux of energy into the sub-filter scales.

Figure \ref{fig:15} gives the kinetic energy dissipation rate metrics for the Tangent filter employed with RF.
The $\epsilon_{S_{ij}}$ metric in Figure \ref{fig:15a} confirms that the chosen tuning only mildly reduces the small-scale activity relative to the reference solution.
Meanwhile the influence of numerical artifacts in the dissipation dynamics is successfully mitigated and driven down upon grid refinement as seen by $\epsilon_{\text{num}}$ in Figure \ref{fig:15b}.
Therefore the attenuation can be said to be properly tuned and sufficiently scale-discriminant in order to ensure that $\epsilon_{num,RF}$ in Equation \ref{eq:38} reduces sufficiently quickly in accordance with the truncation error of the discretization -- thus also highlighting the importance of pairing high accuracy filters with high-order methods.
Interestingly, the overall dissipation rate is in good agreement with the reference solution (see $\epsilon_{E_k}$ in Figure \ref{fig:15c}), despite the lack of an explicit LES model.

Figure \ref{fig:16} includes the kinetic energy dissipation rate metrics for the Tangent filter employed with solution filtering.
Unlike the RF rendition, the active regularization imparted by the SF-r method is seen to reduce the amount of small-scale activity according to $\epsilon_{S_{ij}}$ in Figure \ref{fig:16a}.
Meanwhile, inspecting the dissipation rate due to numerical effects via $\epsilon_{\text{num}}$ in Figure \ref{fig:16b} suggests a growing effect upon grid refinement.
This is directly tied to the filter-induced terms in the kinetic energy equation (see $\epsilon_{num,AD/SF}$ in Equation \ref{eq:36}), whose influence is largely semi-definite and increases in magnitude with $\mathcal{D}_{\text{fil}}$.
Indeed the influence of the regularizing attenuation grows upon grid refinement because the effective filter-to-grid ratio increases accordingly.
This comportment of solution filtering is fundamentally different from that of residual filtering; yet, is seen to also replicate the proper overall dissipation rate in $\epsilon_{E_k}$ in Figure \ref{fig:15c}, despite the lack of an explicit LES model.

\subsection{Contextualizing with respect to Large-Eddy Simulations}  \label{sec:3p4}

The non-linear turbulent-like setting of the Taylor Green vortex test case highlights important nuances between the filtering implementations and their filter pairings -- such as showing different behaviors under grid refinement while holding a constant filter width.
The trends observed relative to the presence of small scale activity (indicated via $\epsilon_{S_{ij}}$) as well as the influence of filter-induced terms in the kinetic energy balance (indicated via $\epsilon_{\text{num}}$) suggest fundamental differences in the filtering methodologies.
Therefore, the respective approaches would likely require different explicit LES closure model pairing strategies.

Indeed an important application of the filtering approaches lies within the context of studying unsteady turbulent flows via LES, which is an accessible alternative to cost-prohibitive Direct Numerical Simulations (DNS).
Because the solution at the grid scales in LES are dynamically active (i.e., exhibit non-trivial spectral energy), the heightened presence of discretization and aliasing errors towards high wavenumbers can impact the accuracy of the overall calculation.
For instance, numerical errors can overwhelm closure model contributions \cite{Ghosal:1996,Moin:1997} and can bias model performance, which typically depends on small scale information. 
Notably, directly discretizing the LES equations on a given grid implies a numerical filter \cite{Geurts:2005} whose properties, in practice, cannot be straightforwardly identified and accounted for via modeling.
In such \emph{implicitly-filtered} LES (IF-LES) methods, the intimate coupling between the implicitly-defined filter and the closure modeling can lead to spurious cancellations of errors for a given grid resolution \cite{Klein:2008} that in turn prevents a steady relaxation towards grid independent results, except at the impractical limit of DNS resolutions.
Explicitly applying a prescribed filter kernel to the LES governing equations attempts to rectify these issues by separating the resolved and unresolved scales with a known filter width that is independent of the grid spacing.
Such \emph{explicitly-filtered} LES (EF-LES) formulations therefore introduce a user-defined resolution scale that can be fixed under grid refinement, which then allows artifacts stemming from the numerical approximation (e.g., discretization and aliasing errors) to be systematically eliminated from the solution\cite{Bose:2010,Bellan:2012,Gallagher:2019,Bertels:2019,Berland:2011}.
As the impacts of numerical errors are no longer conflated with the results, this can then provide a suitable environment for performing objective closure model assessments.
Both residual and solution filtering methodologies have found use in EF-LES implementations -- with the former (RF) focusing on enforcing the spectral support (i.e., a target filter width resolution) of the solution \cite{Lund:2003}, and the latter (SF) more commonly seeking to emulate dissipative effects of traditional closure models \cite{Aubard:2013}.

\section{Conclusions}  \label{sec:4}

Filtering mitigates the presence of small-scale content; however, the means by which this is achieved depends on the specific implementation.
By studying residual and solution filtering from different vantage points (i.e., the equivalent residual equations, a semi-discrete perspective, von Neumann analysis, and numerical computations), the current study highlights the properties associated with the respective methods.
The linear analysis is shown to appropriately characterize the method behaviors in the non-linear settings considered herein.

Inspection of the ERE associated with a prototypical advection-diffusion equation conveys that the RF procedure mollifies the original dynamics by inducing dispersive and anti-diffusive terms, with the associated Von Neumann analysis confirming a general reduction in numerical damping and phase accuracy.
The dispersion effects added by residual filtering yield a lagging behavior and can cause structure incoherence with respect to transport problems.
Yet the dispersive contributions of RF are also responsible for slowing the cascade of energy in non-linear settings -- a notion recently studied by Yalla \emph{et al} \cite{Moser:2021} in terms of the phase errors associated with different schemes. 
Importantly, we note that residual filtering does not actively suppress small-scale content but rather aims to inhibit the accumulation of such content in the first place.
Due to the lack of an inherent damping mechanism to remove high wavenumber noise, the ability to enforce a target filter width in long-time calculations depends heavily on the spectral response of the filter scheme in use.
Scale discriminant filter formulations (e.g., the Tangent filter scheme) slow the generation of high wavenumber content while successfully preserving the intended resolved scales.
By contrast, filters featuring highly smooth responses allow for more small-scale content to gather, and cannot hold the intended filter resolution over long periods of time.

On the other hand, the SF implementation is a dissipative post-processing of the solution and therefore strictly adds regularization without affecting the phase characteristics of the base method.
While solution filtering is able to actively remove small-scale content, it is also sensitive to the filtering scheme.
Employing filters with smooth responses can yield overly damped calculations that inhibit turbulence physics, for example.
As such, properly tuned scale-discriminant filters are also useful in the SF context for keeping the effective filter width from growing in long-time calculations.
Analogous conclusions apply to artificial dissipation methods, which closely resemble the SF procedures in the limit of a sufficient time-step refinement. 

Although residual and solution filtering function via different mechanisms, they can both be applied favorably towards enforcing a target filter width and reducing the presence of small scale content.
With respect to the LES context, however, each implementation will likely entail unique requirements for closure modeling.
For example, RF can be viewed as zeroth-order approximate deconvolution modeling method \cite{Stolz:1999} and therefore could benefit from employing structural models that reconstruct the resolved sub-filter scales.
This may help in terms of recovering the proper non-linear cascade for RF in the presence of non-sharp filters.
Meanwhile, it may be favorable to employ SF adaptively as a model in-itself, likening it to a dynamic hyper viscous eddy-viscosity closure \cite{Lamballais:2021}. 
Such assessments constitute important lines of future research towards the effective use of filtering in the predictive deployment of LES, as well as the objective development and assessment of LES closure models.

\section*{Acknowledgments}

\subsection*{Financial disclosure}

Simulations were supported in part by a grant of computer time from the DOD High Performance Computing Modernization Program.
Funding for this work was supported by the Air Force Office of Scientific Research (AFOSR) (program officer: Dr. Chiping Li), as well as Jacobs Engineering Inc. under contract No. FA9300-20-F-9801 and Innovative Scientific Solutions Inc. under contract No. FA8650-19-F-2046.

\subsection*{Conflict of interest}

The authors declare no potential conflict of interests. Any opinions, findings, and conclusions or recommendations expressed in this material are those of the author(s) and do not necessarily reflect the views of the United States Air Force.

\appendix
\section{Modified Equation Analysis for Artificial Dissipation} \label{app:a}

The following extends the modified equation analysis of artificial dissipation method from previous work \cite{Edoh:2018} to the prototypical advection-diffusion equation.
This is done to highlight the secondary effects that are induced by the spatio-temporal coupling and to clarify the ERE associated with solution filtering.

For simplicity, we assume explicit Euler integration of the 1D advection-diffusion which gives
\begin{eqnarray}
\underbrace{\frac{u^{n+1} - u^n}{\Delta t}}_{\partial_t u^n + \frac{1}{2}(\Delta t) \partial_t^2u + \dots} = - a \delta_x u + \nu \delta_x^2 u^n +  |\lambda'|  \sum_k \epsilon_k (\Delta x)^{2k-1}\delta_x^{2k} u^n  \ .
\end{eqnarray}
The modified equation is derived by substituting the Taylor expansions for the discrete derivatives.
With the intent of focusing on the effect of integration, we look at the leading temporal error term, $\frac{1}{2}(\Delta t)\partial_t^2u$ and re-express it in terms of spatial derivatives:
\begin{eqnarray}
\overbrace{\partial^2_t u}^{\partial_t\{\partial_t u \}}  &=& \left[ - a \partial_x  + \nu \partial_x^2  + \ |\lambda'|  \sum_k \epsilon_k (\Delta x)^{2k-1}\partial_x^{2k} \right] \left\{ - a \partial_x u + \nu \partial_x^2 u + \ |\lambda'|  \sum_k \epsilon_k (\Delta x)^{2k-1}\partial_x^{2k} u\right\}  \\
 &=& a^2 \partial_x^2 u + \nu^2 \partial_x^4 u  + (\lambda')^2\sum_k \epsilon_k (\Delta x)^{4k-2}\partial_x^{4k}u -  2a \nu \partial_x^3u \nonumber \\
 &&   - \ 2a|\lambda'|\sum_k \epsilon_k (\Delta x)^{2k-1}\partial_x^{2k+1}u + 2\nu |\lambda'| \sum_k \epsilon_k (\Delta x)^{2k-1}\partial_x^{2k+2}u \nonumber
\end{eqnarray}
The modified equation is then
\begin{eqnarray}
 \partial_t u &=& -a \partial_x u + \nu \partial_x^2 u + |\lambda'|  \sum_k \epsilon_k (\Delta x)^{2k-1}\partial_x^{2k} u  - \frac{1}{2}(\Delta t) \partial^2_t u + \dots   \\
&=& -a \partial_x u + \nu \partial_x^2 u  \\
&& +   \underbrace{|\lambda'|  \sum_k \epsilon_k (\Delta x)^{2k-1}\partial_x^{2k} u}_{\color{red} I} \nonumber \\
&&  + \underbrace{(\Delta t) \cdot a|\lambda'|\sum_k \epsilon_k (\Delta x)^{2k-1}\partial_x^{2k+1}u}_{\color{blue} II}
   -  \underbrace{(\Delta t) \cdot \nu |\lambda'| \sum_k \epsilon_k (\Delta x)^{2k-1}\partial_x^{2k+2}u}_{\color{cyan} III} + \dots \nonumber
\end{eqnarray}
where the above highlights a pair of induced dispersive and diffusive terms that suggest that the artificial dissipation generates secondary phase and damping effects.
It should be noted that the terms ${\color{blue}II}$ and ${\color{cyan}III}$ are in fact equal yet opposite in sign to the terms found in the ERE of SF (see Equation \ref{eq:8})-- which further confirms the fact that solution filtering only includes the dissipation of the filter to the input with no further secondary effects.

\newpage
\bibliographystyle{aiaa}
\bibliography{./Reference_database.bib}

\end{document}